%% file: BPH-18-005_temp.tex
\pdfoutput=1

\documentclass[11pt,twoside,a4paper,cmspaper,final,collab]{cms-tdr}
\def\svnVersion{dc38a4f-D}\def\svnDate{2019/12/19}\def\cmsCernNoTag{CERN-EP-2019-128}\def\cmsCernDate{\today}\def\cmsMessage{"Published in the Journal of High Energy Physics as \href{http://dx.doi.org/10.1007/JHEP12(2019)100}{\doi{10.1007/JHEP12(2019)100}.}"}

\begin{document}\cmsNoteHeader{BPH-18-005}

\hyphenation{had-ron-i-za-tion}
\hyphenation{cal-or-i-me-ter}
\hyphenation{de-vices}
\RCS$HeadURL$
\RCS$Id$
\ifthenelse{\boolean{cms@external}}{\providecommand{\suppMaterial}{the supplemental material [URL will be inserted by publisher]}}{\providecommand{\suppMaterial}{App.~\ref{app:suppMat}}}

\renewcommand{\arraystretch}{1.1}
\newlength\cmsFigWidth
\ifthenelse{\boolean{cms@external}}{\setlength\cmsFigWidth{0.49\textwidth}}{\setlength\cmsFigWidth{0.65\textwidth}}
\ifthenelse{\boolean{cms@external}}{\providecommand{\cmsLeft}{top\xspace}}{\providecommand{\cmsLeft}{left\xspace}}
\ifthenelse{\boolean{cms@external}}{\providecommand{\cmsRight}{bottom\xspace}}{\providecommand{\cmsRight}{right\xspace}}
\newcommand{\kaon}      {\ensuremath{\PK}\xspace}
\newcommand{\Kstarp}    {\ensuremath{\kaon^{*+}}\xspace}
\newcommand{\Kstarpz}   {\ensuremath{\kaon^{*}(892)^+}\xspace}
\newcommand{\ctK}          {\ensuremath{\cos\theta_{\kaon^*}}\xspace}
\newlength\cmsTabSkip\setlength{\cmsTabSkip}{1ex}

\newcommand{\PP}{\ensuremath{\cmsSymbolFace{P}}\xspace}
\renewcommand{\floatpagefraction}{1.0}
\renewcommand{\topfraction}{1.0}
\renewcommand{\bottomfraction}{1.0}

\cmsNoteHeader{BPH-18-005}
\title{Study of the ${\PBp \to \JPsi\PagL \Pp}$ decay in proton-proton collisions at ${\sqrt{s}= 8\TeV}$}

\date{\today}

\abstract{
A study of the ${\PBp \to \JPsi\PagL \Pp}$ decay using proton-proton collision data collected at $\sqrt{s}= 8\TeV$ by the CMS experiment at the LHC, corresponding to an integrated luminosity of $19.6\fbinv$, is presented. The ratio of branching fractions ${\cal B}(\PBp \to \JPsi\PagL \Pp)/{\cal B}(\PBp \to \JPsi \Kstarpz)$ is measured to be $(1.054 \pm 0.057 \stat \pm 0.035 \syst\pm 0.011({\cal B}))\%$, where the last uncertainty reflects the uncertainties in the world-average branching fractions of $\PagL$ and $\Kstarpz$ decays to reconstructed final states. The invariant mass distributions of the $\JPsi\PagL$, $\JPsi\Pp$, and $\PagL \Pp$ systems produced in the $\PBp \to \JPsi\PagL \Pp$ decay are investigated and found to be inconsistent with the pure phase space hypothesis. The analysis is extended by using a model-independent angular amplitude analysis, which shows that the observed invariant mass distributions are consistent with the contributions from excited kaons decaying to the $\PagL \Pp$ system.
}
\hypersetup{%
pdfauthor={CMS Collaboration},%
pdftitle={Study of the B+ To J/psi Lambda-bar p decay in proton-proton collisions at sqrt(s) = 8 TeV},%
pdfsubject={CMS},%
pdfkeywords={CMS, b physics, heavy-flavor spectroscopy, experimental results}
}

\maketitle
\section{Introduction}

The $\PBp \to \JPsi\PagL \Pp$ decay is the first observed example of a $\PB$ meson decay into baryons and a charmonium state (the charge-conjugate states are implied throughout the paper).
The first evidence for the $\PBp \to \JPsi\PagL \Pp$ decay was obtained by the BaBar Collaboration~\cite{babar_first}, along with a measurement of its branching fraction of $\mathcal{B}(\PBp \to \JPsi\PagL \Pp)=(12^{+9}_{-6})\times10^{-6}$. Later, this decay was observed by the Belle Collaboration and
its branching fraction was measured to be $\mathcal{B}(\PBp \to \JPsi\PagL \Pp) = (11.7\pm2.8^{+1.8}_{-2.3})\times10^{-6}$~\cite{jpsilamp}, resulting in the world-average branching fraction of $\mathcal{B}(\PBp \to \JPsi\PagL \Pp) = (11.8\pm3.1)\times10^{-6}$~\cite{PDG}.

The decay mode under study provides an opportunity to search for new intermediate resonances in the $\JPsi\PagL$, $\JPsi \Pp$, and $\PagL \Pp$ systems. The interest in $\PB$ hadron decays to charmonium-baryon systems has increased since the observation of three pentaquark states in ${\PgL}_{\PQb}^{0} \to \JPsi \Pp \PKm$ decays by the LHCb Collaboration~\cite{lhcb_penta, lhcb_penta_new}.
While these states are beyond the available phase space of the $\PBp \to \JPsi\PagL \Pp$ decay, this process can target potential low-mass pentaquark states in the $\JPsi \Pp$ system, as well as new resonances in the $\JPsi\PagL$ system, where one of the new states is expected to appear close to the threshold and represent itself as a threshold enhancement~\cite{JpsiLambdaTheory}. Recently, the existence of a molecular baryon state decaying to $\JPsi\PagL$ and potentially accessible via the $\PBp \to \JPsi\PagL \Pp$ decay has been predicted~\cite{JpsiLambdaTheoryBM}.

In this paper, we report on a study of the $\PBp \to \JPsi\PagL \Pp$ ($\JPsi \to \PGmp\PGmm$, $\PagL \to \Pap\Pgpp$) decay using a data sample of proton-proton ($\Pp\Pp$) collisions collected by the CMS experiment in 2012 at $\sqrt{s} =8\TeV$, corresponding to an integrated luminosity of 19.6$\fbinv$. Exploring the large available integrated luminosities of $\Pp\Pp$ collisions and the large  production cross  section  of $\bbbar$  pairs at  the  CERN  LHC, the CMS experiment has developed an efficient trigger for displaced $\JPsi \to \PGmp\PGmm$ decays, described in Section~\ref{Selection and Samples}. This trigger allowed CMS to conduct this study of the $\PBp \to \JPsi\PagL \Pp$ decay, including the measurement of its branching fraction and the study of the $\JPsi\PagL$, $\JPsi \Pp$, and $\PagL \Pp$ systems. The decay $\PBp \to \JPsi \Kstarpz$ ($\Kstarpz  \to \PKzS\Pgpp \to \Pgpp \Pgpm\Pgpp $) is chosen as the normalization channel, because it is measured with high precision and has a similar decay topology to the $\PBp \to \JPsi\PagL \Pp$ decay. In what follows, the $\Kstarpz$ particle is denoted as $\Kstarp$. The ratio of the branching fractions is measured using the following formula:
\begin{linenomath}
\begin{equation}\label{BFeq}
\dfrac{\mathcal{B}(\PBp \to \JPsi\PagL \Pp)}{\mathcal{B}(\PBp \to \JPsi \Kstarp)}
 = \dfrac{N(\PBp \to \JPsi\PagL \Pp)\mathcal{B}(\Kstarp  \to \PKzS\Pgpp)\mathcal{B}(\PKzS \to\Pgpp\Pgpm)\epsilon(\PBp \to \JPsi \Kstarp)}{N(\PBp \to \JPsi \Kstarp)\mathcal{B}(\PagL \to \Pap \Pgpp)\epsilon(\PBp \to \JPsi\PagL \Pp)},
\end{equation}
\end{linenomath}
where $N$ and $\epsilon$ correspond to the number of observed decays and the total efficiency of the decay, respectively. The total efficiency includes the product of efficiencies for the subsequent decays $\Kstarp  \to \PKzS\Pgpp$, $\PKzS \to\Pgpp\Pgpm$, and $\PagL \to \Pap \Pgpp$.
The invariant mass distributions of the $\JPsi\PagL$, $\JPsi\Pp$, and $\PagL \Pp$ systems
produced in the $\PBp \to \JPsi\PagL \Pp$ decay are investigated using a model-independent angular analysis.

\section{The CMS detector}

The central feature of the CMS apparatus is a superconducting solenoid of 6$\unit{m}$ internal diameter, providing a magnetic field of 3.8$\unit{T}$. Within the solenoid volume are a silicon pixel and strip tracker, a lead tungstate crystal electromagnetic calorimeter, and a brass and scintillator hadron calorimeter, each composed of a barrel and two endcap sections. Forward calorimeters extend the pseudorapidity ($\eta$) coverage provided by the barrel and endcap detectors. Muons are detected in gas-ionization chambers embedded in the steel flux-return yoke outside the solenoid. The main subdetectors used for the present analysis are the silicon tracker and the muon system.

The silicon tracker measures charged particles within the range $\abs{\eta} < 2.5$. During the LHC running period when the data used in this paper were recorded, the silicon tracker consisted  of 1440 silicon pixel and 15\,148 silicon strip detector modules. The track resolutions are typically 1.5$\%$ in transverse momentum ($\pt$) and 25--90 (45--150)$\mum$ in the transverse (longitudinal) impact parameter~\cite{d1} for nonisolated particles with $1 < \pt < 10\GeV$ and $\abs{\eta} < 1.4$.

Muons are measured within $\abs{\eta} < 2.4$, with detection planes made using three technologies: drift tubes, cathode strip chambers, and resistive-plate chambers. Matching muons to tracks measured in the silicon tracker results in a relative $\pt$ resolution of 0.8--3.0$\%$ for muons with $\pt < 10\GeV$ used in this analysis, depending on the muon $\abs{\eta}$~\cite{d2}.

Events of interest are selected using a two-tiered trigger system~\cite{Khachatryan:2016bia}. The first level (L1), composed of custom hardware processors, uses information from the calorimeters and muon detectors to select events at a rate of around 100$\unit{kHz}$ within a time interval of less than 4$\mus$. The second level, known as the high-level trigger (HLT), consists of a farm of processors running a version of the full event reconstruction software optimized for fast processing, and reduces the event rate to around 1$\unit{kHz}$ before data storage.

A more detailed description of the CMS detector, together with a definition of the coordinate system used and the relevant kinematic variables, can be found in Ref.~\cite{d3}.

\section{Data sample and event selection} \label{Selection and Samples}

Data were collected with a dedicated trigger, optimized for the selection of $\PQb$ hadrons decaying to $\JPsi(\PGmp\PGmm)$.  The L1 trigger required two oppositely charged muons, each with
$\pt> 3\GeV$ and $\abs{\eta}<2.1$. At the HLT, a $\JPsi$ candidate decaying into a $\PGmp\PGmm$ pair displaced from the interaction point was required. Each muon must have $\pt>4\GeV$, and the dimuon $\pt$ must exceed $6.9\GeV$. The HLT demanded that $\JPsi$ candidates reconstructed from opposite-sign dimuons have an invariant mass between 2.9 and 3.3\GeV. The three-dimensional (3D) distance of
closest approach of the two muons of a pair to each other was required to be less than 0.5\unit{cm}. The dimuon vertex fit was required to have a transverse decay length significance
$L_{xy}/\sigma_{L_{xy}} > 3$, where $L_{xy}$ and $\sigma_{L_{xy}}$
are, respectively, the distance from the common vertex to the beam axis in the transverse plane, and its uncertainty.
Finally, the dimuon vertex fit probability, calculated using the
$\chi^2$ and the number of degrees of freedom of the vertex fit, was required to exceed 10$\%$, while
the angle $\alpha$ between the dimuon $\pt$ vector and the
direction connecting the beam spot and the dimuon vertex in the transverse plane was required to satisfy $\cos\alpha>0.9$.

The analysis requires two muons of opposite charge that must match those that triggered the event readout. The trigger requirements are confirmed and the $\JPsi$ candidates are selected by tightening the dimuon mass region to be within 150$\MeV$ of the $\JPsi$ meson mass $M_\JPsi^\text{PDG}$~\cite{PDG} ($M_\text{X}^\text{PDG}$ denotes the world-average mass of hadron X).

To reconstruct a $\PBp$ candidate, the $\JPsi$ candidate is combined with a positively charged particle track, assumed to be a proton track, and a $\PagL$ candidate. The track must satisfy the CMS high-purity requirements~\cite{d1}. The $\PagL$ candidates are formed from displaced two-prong vertices under the assumption of the $\PagL  \to \Pap \Pgpp$ decay, as described in Ref.~\cite{trackreco}. Daughter particles of the $\PagL$ candidate are refitted to a common vertex, and the vertex fit probability must exceed $1\%$. The proton mass is assigned to the higher-$\pt$ daughter track. To select the candidates in the $\PagL$ signal region, we demand that the $\Pap\Pgpp$ invariant mass satisfy $\abs{M(\Pap\Pgpp)-M_{\PagL}^\text{PDG}}<2\sigma^{\PagL}$, where the effective $\PagL$ signal resolution $\sigma^{\PagL} = 3.7\MeV$ is measured in data by fitting the $M(\Pap \Pgpp)$ distribution with a sum of two Gaussian functions with a common mean.

Since reliable charged hadron type identification in CMS is not possible, the contribution from $\PKzS \to \Pgpm\Pgpp$ decays is present in the $\PagL  \to \Pap \Pgpp$ sample. It is removed by demanding that the invariant mass of the two $\PagL$ candidate tracks, where both are assigned the charged pion mass, satisfies $\abs{M(\Pgpp\Pgpm)-M_{\PKzS}^\text{PDG}}>2\sigma^{\PKzS}$, where the effective $\PKzS$ signal resolution $\sigma^{\PKzS} = 9.2\MeV$ is measured in data by fitting the $M(\Pgpp \Pgpm)$ distribution with a sum of two Gaussian functions with a common mean.

As the last step of the reconstruction, the kinematic vertex fit of the $\PagL$ candidate, the proton track, and the dimuon is performed, with the dimuon mass constrained to $M_\JPsi^\text{PDG}$; this vertex is referred to as the $\PB$ vertex.
The selected candidates are required to have $\pt(\JPsi)>7\GeV$, $\pt(\PagL)>1\GeV$, and $\pt(\Pp)>1\GeV$.

Multiple $\Pp\Pp$ interactions in the same or nearby beam crossing (pileup) are present in data, with an average multiplicity of about 20.
The hard-scattering vertex in the event with the highest cosine of the three-dimensional (3D) pointing angle between the line connecting this vertex with the $\PB$ vertex and the $\PBp$ candidate momentum is chosen as the primary vertex (PV).  The following requirement is used to select $\PBp$ candidates consistent with originating from the PV: $\cos\alpha (\PBp,\text{PV})>0.99$, where $\alpha(\PBp,\text{PV})$ is the two-dimensional (2D) angle in the transverse plane between the $\PBp$ candidate momentum and the vector pointing from the PV to the $\PB$ vertex. The following requirement on the $\PB$ vertex displacement is also applied: $L_{xy}(\PBp)/\sigma_{L_{xy}(\PBp)} >3$, where $L_{xy}(\PBp)$ is the distance between the primary and $\PB$ vertices in the transverse plane, and $\sigma_{L_{xy}(\PBp)}$ is its uncertainty. The $\PBp$ candidate kinematic vertex fit probability must exceed $1\%$. The $\PagL$ candidate is required to be consistent with originating from the $\PBp$ decay by requiring $\cos\alpha({\PagL,\PBp})>0$, where $\alpha({\PagL,\PBp})$ is the angle between the $\PagL$ momentum and the vector connecting the $\PBp$ and $\PagL$ vertices.

The normalization decay channel $\PBp \to \JPsi \Kstarp$ ($\Kstarp  \to \PKzS\Pgpp \to \Pgpp \Pgpm\Pgpp $) candidates are selected using the same reconstruction chain. Identical requirements are used to select the $\JPsi$ candidate and the $\Pgpp$ track from the $\Kstarp$ meson decay.
The selection of the $\PKzS$ candidates is the same as for the $\PagL$ candidates, except that the invariant mass of the candidate pions from the $\PKzS$ meson decay is required to satisfy $\abs{M(\Pgpp\Pgpm)-M_{\PKzS}^\text{PDG}}<2\sigma^{\PKzS}_\text{eff}$.  The contamination of the $\PKzS$ candidates from the $\PagL  \to \Pap \Pgpp$ decay is reduced with the requirement: $\abs{M(\Pap\Pgpp)-M^\text{PDG}_{\PagL}} >2\sigma^{\PagL}_\text{eff}$, where the negatively charged track is assigned the proton mass.

To calculate the reconstruction efficiency, a study based on simulated signal events is performed. The events are generated with $\PYTHIA$ v6.424~\cite{pythia}. The $\PB$ meson decays are modeled according to a phase space decay model using $\EVTGEN$ v1.3.0~\cite{evtgen} for both the $\PBp \to \JPsi\PagL \Pp$ and $\PBp \to \JPsi \Kstarp$ channels. The simulation sample corresponding to the $\PBp \to \JPsi \Kstarp$ decay is reweighted according to the angular distributions of the $\Kstarp$ and $\JPsi$ systems observed in data by applying a weight to each simulated event obtained using a linear interpolation of the data-to-simulation ratio histogram for each angular variable. The events are passed through a detailed CMS detector simulation based on $\GEANTfour$~\cite{geant4}. To estimate reconstruction efficiencies (Section~\ref{EffSec}), matching of the reconstructed candidates to the generated particles is performed by requiring  $\Delta R = \sqrt{\smash[b]{(\Delta\eta)^2+(\Delta\phi)^2}} <0.004$ (0.03) for $\PGm$ ($\Pp$, $\Pgpp$, $\PagL$, and $\PKzS$) candidates, where $\Delta\eta$ and $\Delta\phi$ are the differences in the pseudorapidity and the azimuthal angle, respectively, between the momenta of the reconstructed and generated particles.

\section{Signal yields extraction}
\begin{figure}[htb]
\centering
\includegraphics[width=0.49\textwidth]{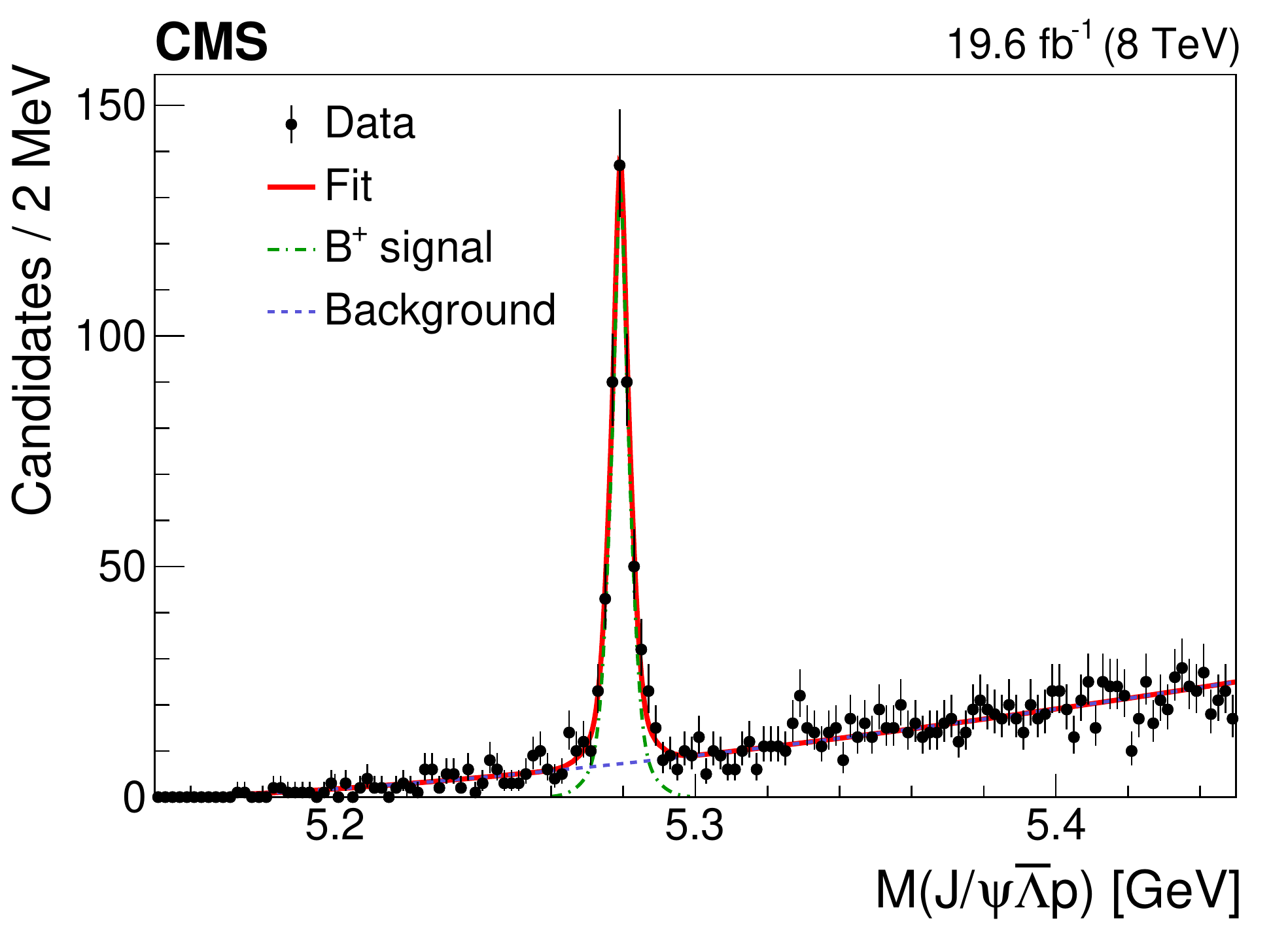}\\
\includegraphics[width=0.49\textwidth]{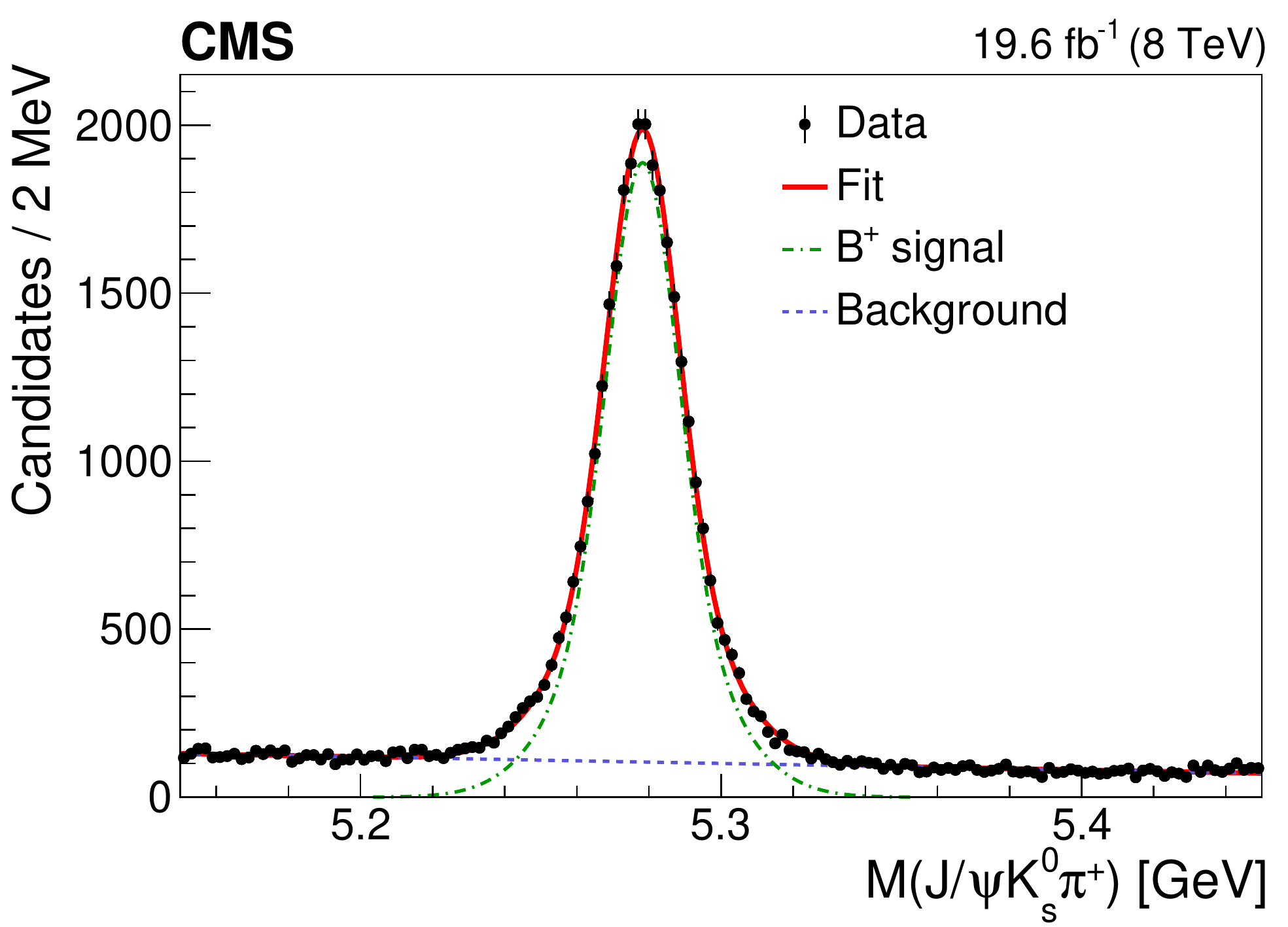}
\includegraphics[width=0.49\textwidth]{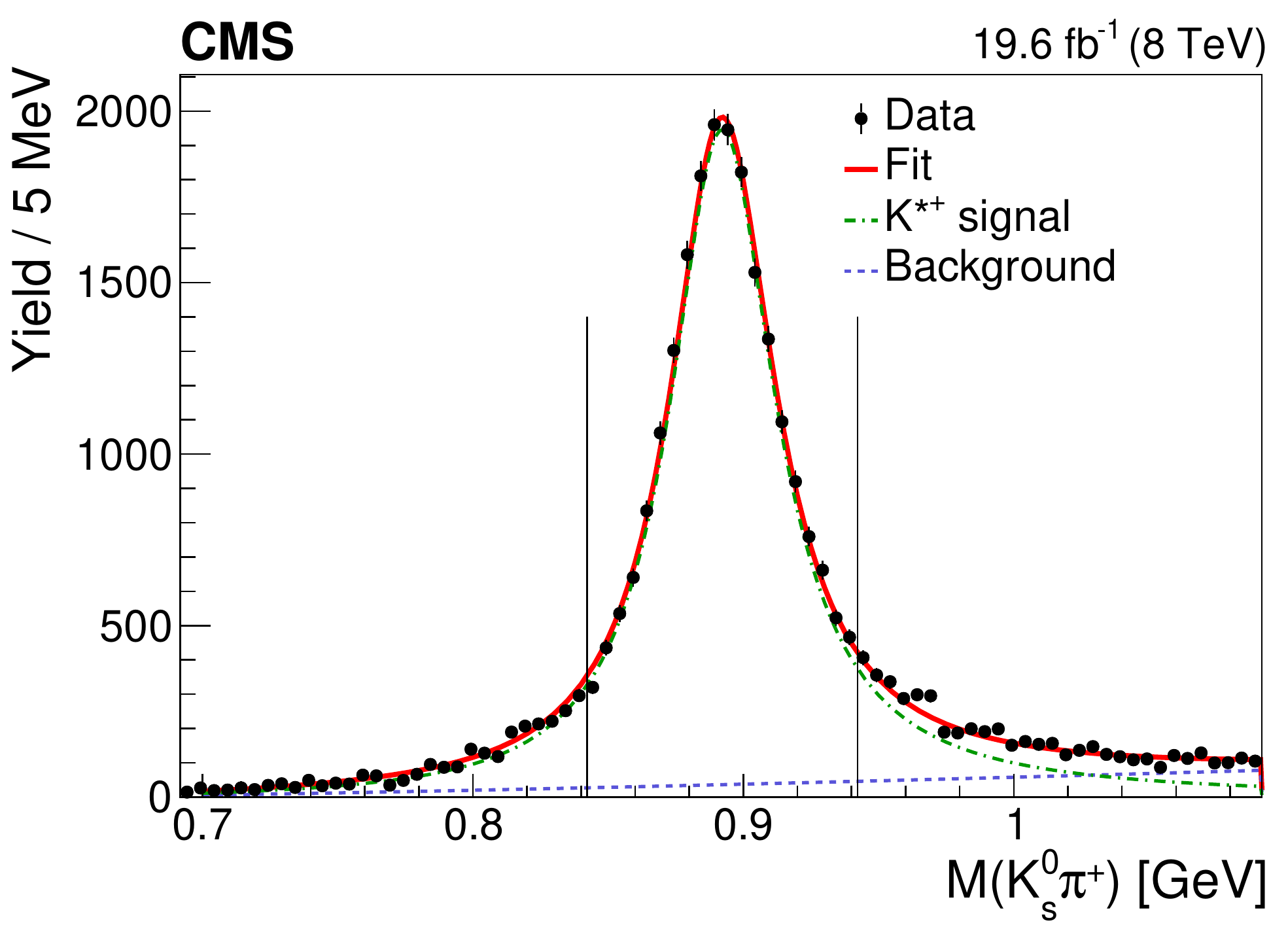}
\caption{
The invariant mass distribution of the selected $\PBp \to \JPsi\PagL \Pp$ candidates (upper).
The invariant mass distributions of $\JPsi \PKzS\Pgpp$ (lower left) and $\PKzS\Pgpp$ (lower right) for the $\PBp \to \JPsi \Kstarp$ decay candidates. The points are data and the solid curves are the results of the fits explained in the text. The vertical bars represent the statistical uncertainty. On the lower right picture the background-subtracted candidates using the $M(\JPsi \PKzS\Pgpp)$ as a discriminating variable are shown. The dash-dotted curves show the $\PBp$ signal in the upper and lower left plots, and the $\Kstarp$ signal in the lower right plot. The dashed lines indicate the background contributions. The vertical lines in the lower right plot indicate the $\Kstarp$ invariant mass window used for the normalization, as described in the text.}
\label{m2}
\end{figure}

The invariant mass distribution of the selected $\PBp \to \JPsi\PagL \Pp$ candidates is shown in Fig.~\ref{m2} (upper). An unbinned, extended maximum-likelihood fit with a signal plus background hypothesis is performed on this distribution. The signal component is modeled with a sum of three Gaussian functions with floating common mean and overall normalization, while the widths and the relative normalizations of the three Gaussian functions are fixed to the values obtained from simulation. The background component is parameterized by a threshold function: $(x-x_{0})^{\beta}$, where $x_{0} = M_{\PagL}^\text{PDG} + M_{\JPsi}^\text{PDG} +  M_{\Pp}^\text{PDG}$ and $\beta$ is a free parameter of the fit. The fit results in a signal yield of $452\pm 23$ events.

Figure~\ref{m2} (lower left) shows the observed $\PBp \to \JPsi \PKzS\Pgpp$ invariant mass distribution with the requirement on the $\PKzS\Pgpp$ invariant mass to be inside a $\pm200\MeV$ window around the world-average $\Kstarp$ mass~\cite{PDG}.
The $\PBp$ signal is modeled with a sum of two Gaussian functions with a common mean and all the parameters floating in the fit, while the background is described by a second-order polynomial.

In order to evaluate the pure $\Kstarp$ meson contribution, excluding other resonances in the $\PKzS\Pgpp$ system, the background-subtracted $M(\PKzS\Pgpp)$ distribution shown in Fig.~\ref{m2} (lower right) is fitted in the range of $\pm200\MeV$ around the $\Kstarp$ mass. Background subtraction is performed using the \textit{sPlot} technique~\cite{sPlot} with the $M(\JPsi \PKzS\Pgpp)$ used as the discriminating variable. The instrumental mass resolution of the $\Kstarp$ peak is negligible in comparison with its natural width; therefore a relativistic Breit--Wigner function is used as the signal model, while a threshold polynomial function is chosen to model the non-$\Kstarp$ component: $(x-x_{0}^{\kaon^{*}})^{\gamma}$, where $x_{0}^{\kaon^{*}} = M_{\PKzS}^\text{PDG} + M_{\Pgpp}^\text{PDG}$ and $\gamma$ is a free parameter in the fit.

To obtain the observed number of $\PBp \to \JPsi \Kstarp$ decays for the measurement of the ratio of branching fractions, the signal Breit--Wigner function is integrated over $\pm50\MeV$ around the $\Kstarp$ mass, resulting in an yield of $20~863\pm357$ events. The efficiency of the requirement on $M(\PKzS\Pgpp)$ of being within $\pm50\MeV$ of the $\Kstarp$ mass is taken into account in the calculation of the total efficiency (Section~\ref{EffSec}).

\section{Efficiency calculation}\label{EffSec}

The efficiency for detecting and identifying the $\PBp \to \JPsi\PagL \Pp$ decay is calculated as the ratio of the numbers of reconstructed to generated events in simulation. The overall efficiency includes the trigger and reconstruction efficiencies, and the detector acceptance. The efficiency in each channel is obtained using simulated samples described in Section~\ref{Selection and Samples}. The efficiency ratio, which is used in the measurement of the ratio of branching fractions, is found to be $\epsilon(\PBp \to \JPsi \Kstarp)/\epsilon(\PBp \to \JPsi\PagL \Pp) = 1.347 \pm 0.023$, where the uncertainty is statistical only and accounts for the limited event counts in the corresponding simulated samples.

\begin{figure}[htb]
\centering
\includegraphics[width=0.9\textwidth]{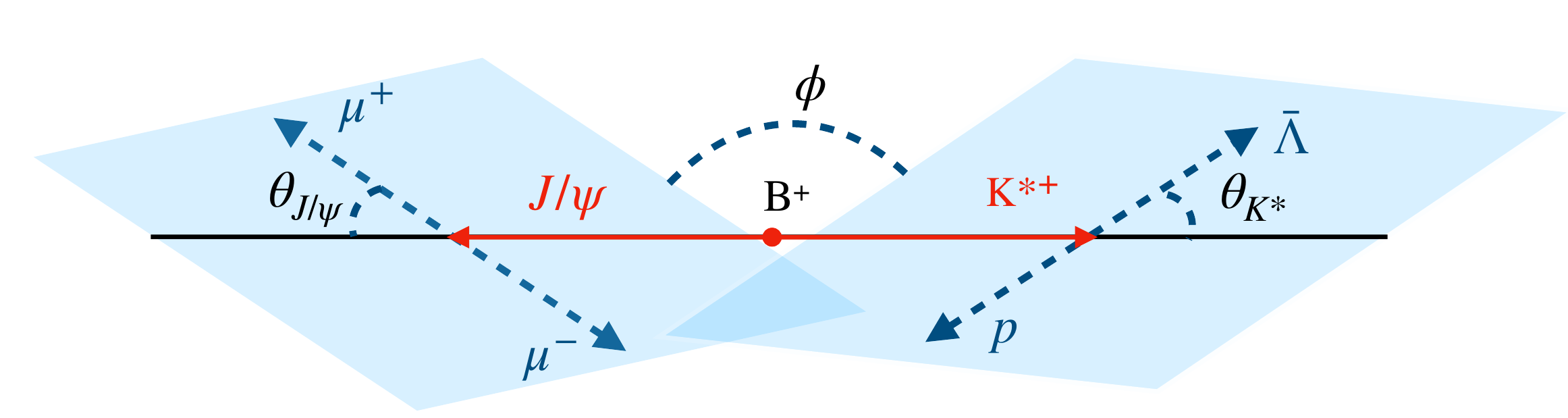}
\caption{
An illustration of the decay angles in the $\PBp \to \JPsi\Kstarp_{2,3,4}(\PagL \Pp)$ decay.}
\label{DecAngl}
\end{figure}

For the study of the two-body intermediate invariant masses in the $\PBp \to \JPsi\PagL \Pp$ decay, we perform an efficiency-correction procedure to account for detector effects. Crucial to the investigation of the $\PagL\Pp$ system is the possibility of intermediate high-mass $\Kstarp$ resonances that can decay to $\PagL \Pp$~\cite{PDG}. We list these resonances in Table~\ref{Kstars} and designate them collectively as $\Kstarp_{2,3,4}$. The details of the decay $\PBp \to \JPsi \Kstarp_{2,3,4}$, followed by $\Kstarp_{2,3,4} \to \PagL\Pp$ are discussed in Section~\ref{invMassStudy}. Because of this possibility, the efficiency is calculated as a function of two variables: the invariant mass of the $\PagL\Pp$ system, $M(\PagL\Pp)$, and the cosine of the $\Kstarp_{2,3,4} \to \PagL\Pp$ system helicity angle $\ctK$, which is defined as the angle between the $\PagL$ and $\PBp$ momentum vectors in the $\PagL \Pp$ system rest frame (as illustrated in Fig.~\ref{DecAngl} together with other decay angles). The 2D efficiency is calculated as the ratio of the 2D histogram at the reconstruction level to that at the generator level. The data are corrected for the reconstruction efficiency by applying a $1/\epsilon(M(\PagL\Pp),\ctK)$ weight to each event. Efficiency values at each point in the 2D space are evaluated using a bilinear interpolation algorithm. Since the points inside the border bins of the 2D space cannot be interpolated, the efficiency values at these points are assumed to be the values at the centers of the corresponding bins.

\begin{table}[htb]
\centering
\topcaption{The mass, width, and $\text{J}^{\text{P}}$ quantum numbers for the known $\Kstarp$ states~\cite{PDG} that can decay to $\PagL\Pp$.}
\begin{tabular}{lccr}
 Resonance  & Mass ($\MeV$) & Natural width ($\MeV$) & $\text{J}^{\text{P}}$ \\
 \hline
   $\PKstiv{}^+$                  & $2045\pm 9$   & $198\pm 30$   & $4^{+}$  \\
   ${\PK^*_2(2250)}{}^+$  & $2247\pm 17$  & $180 \pm 30$  & $2^{-}$  \\
  ${\PK^*_3(2320)}{}^+$   & $2324\pm 24$  & $150\pm 30$   & $3^{+}$  \\
\end{tabular}
\label{Kstars}
\end{table}

\section{Evaluation of the systematic uncertainties in the branching fraction ratio measurement}\label{systematic}

In this section, we discuss the sources of the systematic uncertainty in the measurement of the ratio ${\cal B}(\PBp \to \JPsi\PagL \Pp)/{\cal B}(\PBp \to \JPsi \kaon^{*+})$, defined by Eq.~(\ref{BFeq}).

Since the signal $\PBp \to \JPsi\PagL \Pp$ and the normalization $\PBp \to \JPsi \kaon^{*+}$ decays have the same topology, the systematic uncertainties related to the muon reconstruction, track reconstruction, and trigger efficiencies should almost cancel out in Eq.~(\ref{BFeq}). To check if this is the case, simulated signal samples for both $\PBp \to \JPsi\PagL \Pp$ and $\PBp \to \JPsi \kaon^{*+}$ decays are validated by comparing distributions of variables used in the event selection between background-subtracted data and simulated signal samples. As a result of these studies, an additional systematic uncertainty is assigned to account for the deviation between data and simulation in the $\eta$ distributions of $\PBp$ meson for the signal and normalization channels, and the $M(\PKzS\Pgpp)$ distribution for the $\PBp \to \JPsi \kaon^{*+}$ decay. The deviation in the $\eta$ distributions of $\PBp$ meson were taken into account by reweighting the simulated samples according to the $\eta$ distributions of $\PBp$ meson observed on data and recalculating the efficiency ratio using the reweighted simulation samples; the systematic uncertainty is calculated as the difference in efficiency ratios calculated before and after reweighting. The difference in the $M(\PKzS\Pgpp)$ distribution is taken into account by altering the baseline mass window width used for $\kaon^{*+}$ selection of 50\MeV, with 35 and 70\MeV windows, and recalculating the final branching fraction value; the largest deviation from the value obtained using the baseline selection is considered as systematic uncertainty.

The systematic uncertainty related to the choice of the background model is estimated separately for the fits to the $\JPsi\PagL \Pp$, $\JPsi \PKzS\Pgpp{}$, and $\PKzS\Pgpp$ invariant mass distributions.
To evaluate this uncertainty, several additional background modeling functions were used: polynomials of the first, second, and third order for the $\JPsi \PKzS\Pgpp$ invariant mass distribution; and the function $(x-x_{0})^{\delta}$ multiplied by polynomials of the first and second order for the $\JPsi\PagL \Pp$ and $\PKzS\Pgpp$ invariant mass distributions.

Another source of systematic uncertainty is due to the modeling of the signal shape in the $M(\JPsi\PagL \Pp)$ and $M(\JPsi \PKzS\Pgpp)$ invariant mass distributions. In the case of the $\PBp \to \JPsi\PagL \Pp$ decay, the resolution functions in the baseline fits are obtained from simulation. The associated uncertainty is estimated by allowing the widths to float in the fit and by adding a double-Gaussian function as a fit option.
For the $\PBp \to \JPsi \kaon^{*+}$ decay, the corresponding systematic uncertainty is estimated by using alternative signal models to fit the $\JPsi \PKzS\Pgpp$ invariant mass distribution, such as a sum of three Gaussian functions or two Crystal Ball functions~\cite{CB1,CB2}.

For each of the variations, the largest deviation in the measured signal yield is used as the systematic uncertainty. Variations of the signal and background components are performed independently. The uncertainty in the relative efficiency from the simulation related to the limited number of events in the simulated samples is considered as an additional source of systematic uncertainty.

Table~\ref{table_systematics} summarizes the individual systematic uncertainties, as well as the total systematic uncertainty, calculated as the quadratic sum of the individual components.

\begin{table}[htb]
\centering
\topcaption{Summary of the relative systematic uncertainties in the ${\cal B}(\PBp \to \JPsi\PagL \Pp)/{\cal B}(\PBp \to \JPsi \Kstarp)$ ratio.}
\begin{tabular}{lr}
Source  &  Relative uncertainty ($\%$)\\
\hline
Discrepancy between data and simulation & 2.2\\
Background model in the $M(\JPsi\PagL \Pp)$ distribution&1.1\\
Background model in the $M(\JPsi \PKzS\Pgpp)$ distribution& 0.1 \\
Background model in the $M(\PKzS\Pgpp)$ distribution  &1.2\\
Signal model in the $M(\JPsi\PagL \Pp)$ distribution& 0.9\\
Signal model in the $M(\JPsi \PKzS\Pgpp$) distribution & 0.6\\
Simulated sample event count & 1.7\\[\cmsTabSkip]
Total systematic uncertainty & 3.3 \\
\end{tabular}
\label{table_systematics}
\end{table}

\section{Measurement of the ratio of branching fractions}

{\tolerance=800
Using the world-average values of the $\mathcal{B}(\Kstarp\to\PKzS\Pgpp)$, $\mathcal{B}(\PKzS \to\Pgpp\Pgpm)$, $\mathcal{B}(\PagL  \to \Pap\Pgpp)$ branching fractions~\cite{PDG},
the relative efficiency described in Section~\ref{EffSec},
and the signal yields, the ratio
${\cal B}(\PBp \to \JPsi\PagL \Pp)/{\cal B}(\PBp \to \JPsi \Kstarp)$ is measured using Eq.~(\ref{BFeq}) to be $(1.054\pm0.057\stat \pm0.035\syst\pm0.011({\cal B}))\%$, where the first uncertainty is statistical, the second is systematic (as discussed in Section~\ref{systematic}), and the third is due to the uncertainties in the world-average branching fractions of the decays involved.

From this ratio and the world-average value of ${\cal B}(\PBp \to \JPsi \Kstarp)=(1.43\pm0.08)\times 10^{-3}$~\cite{PDG}, the branching fraction ${\cal B}(\PBp \to \JPsi\PagL \Pp)$ = $(15.1\pm0.8\stat \pm0.5\syst\pm0.9({\cal B}))\times 10^{-6}$ is obtained, where now the last uncertainty includes the uncertainty in the $\PBp \to \JPsi \Kstarp$ branching fraction. This measurement is the most precise to date.
\par}

\section{Study of two-body invariant mass spectra}\label{invMassStudy}

\begin{figure}[htb]
\centering
\includegraphics[width=0.49\textwidth]{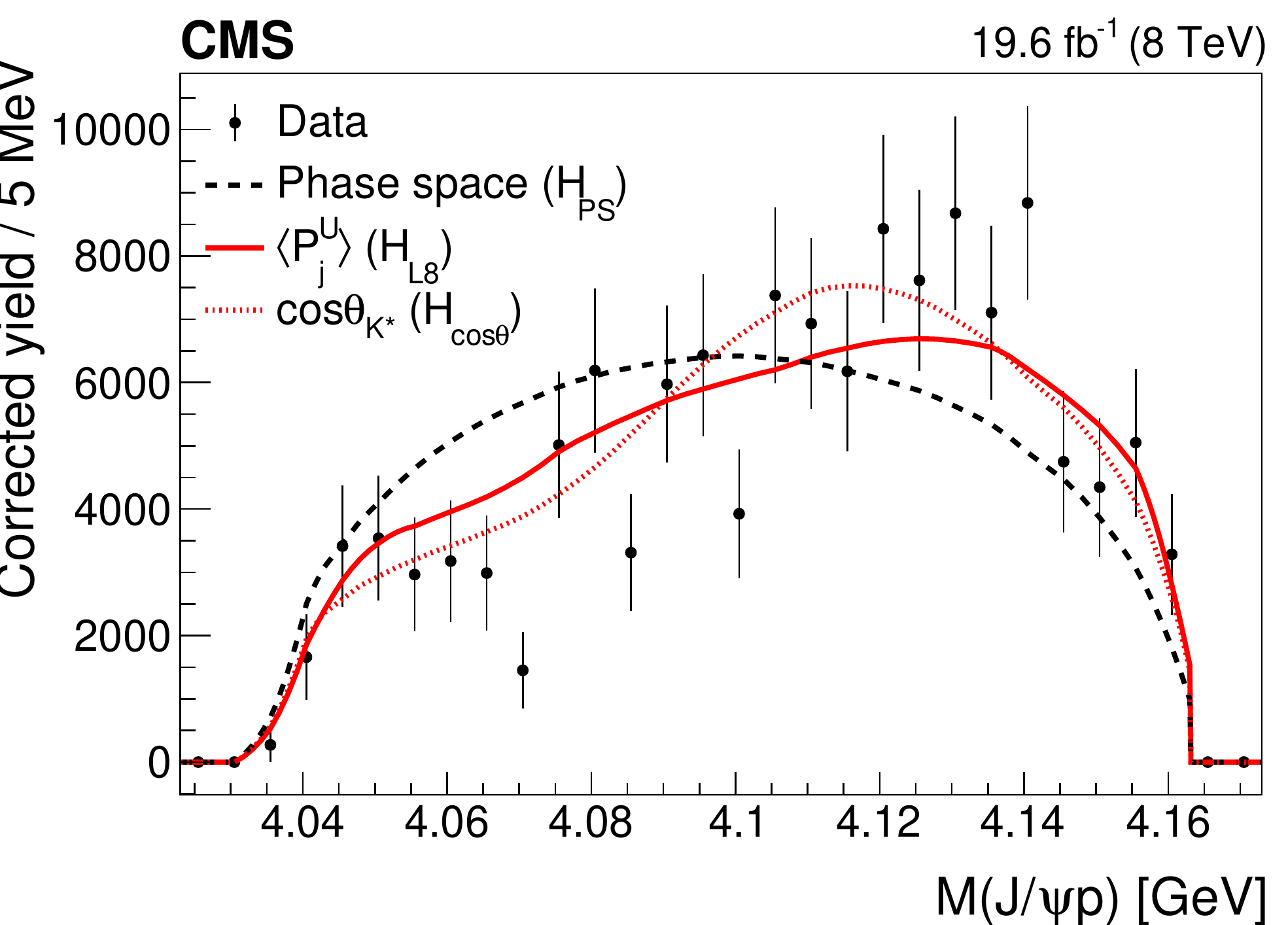}
\includegraphics[width=0.49\textwidth]{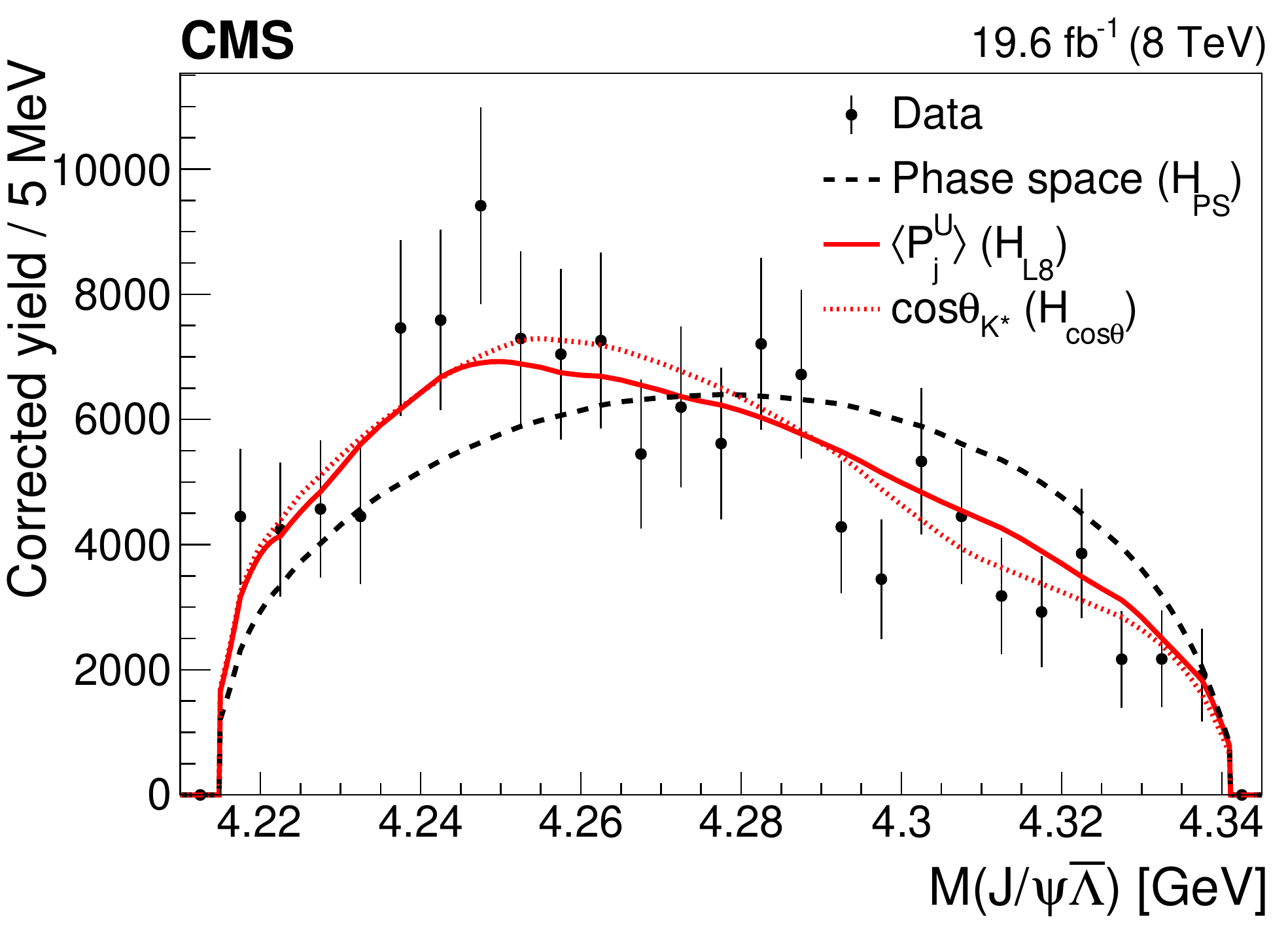}
\includegraphics[width=0.49\textwidth]{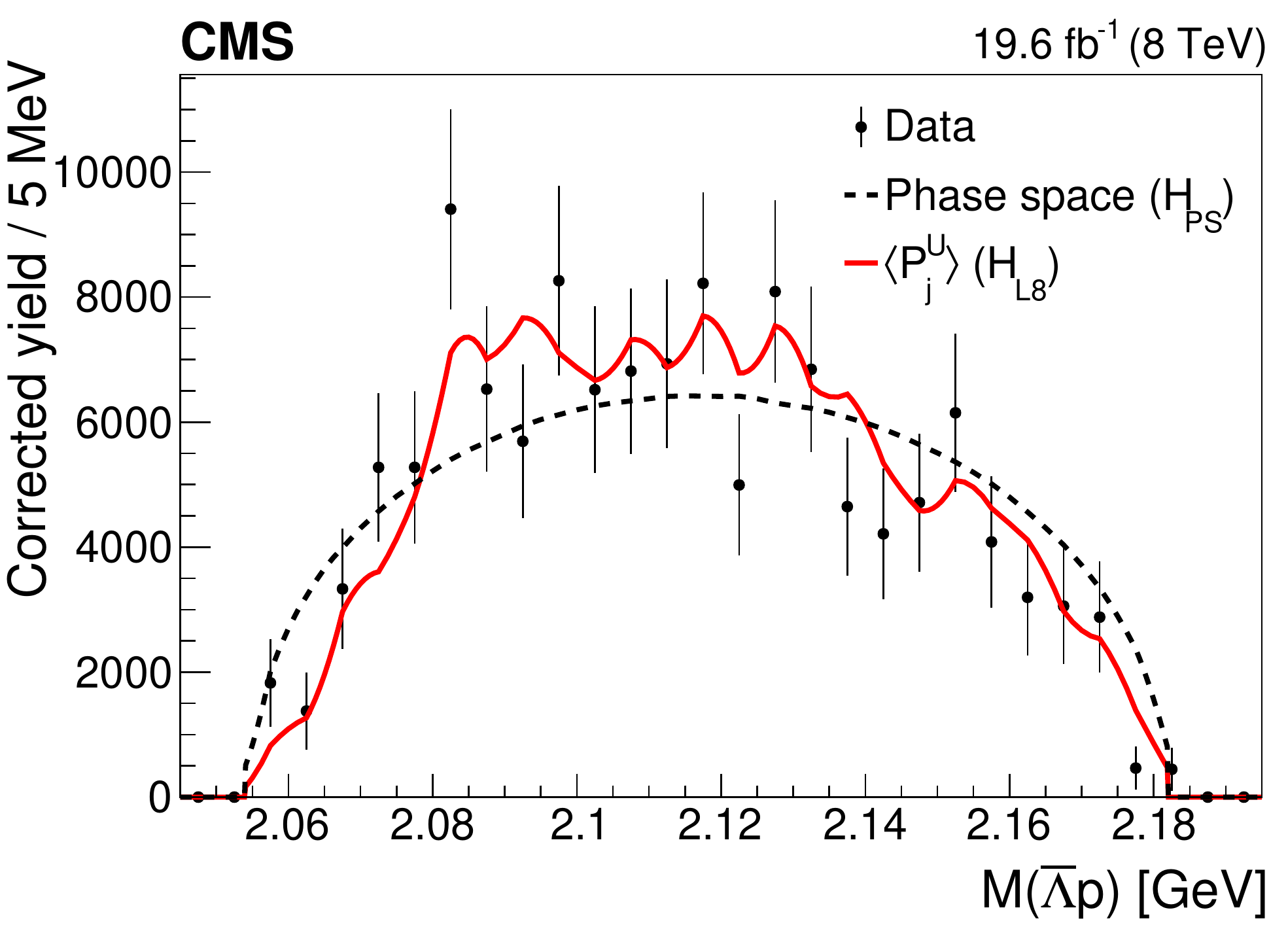}
\caption{
The invariant mass distributions of the $\JPsi\Pp$ (upper left), $\JPsi \PagL$ (upper right), and $\PagL \Pp$ (lower) systems from the $\PBp \to \JPsi\PagL \Pp$ decay.
The points show the efficiency-corrected, background-subtracted data; the vertical bars represent the statistical uncertainty. Superimposed curves are obtained from simulation: the dashed lines correspond to the pure phase space distribution ($H_{\text{PS}}$); the solid curves represent the phase space distribution corrected for the $\PagL \Pp$ angular structure with the inclusion of the first eight moments, corresponding to resonances decaying to the $\PagL\Pp$ system with maximum spin of 4 ($H_{\text{L8}}$); the dotted curves show the phase space distribution reweighted according to the $\ctK$ distribution, which is defined as the $H_{\cos\theta}$ hypothesis. The mentioned curves are explained in Section~\protect\ref{significance}.}
\label{m12}
\end{figure}

In this section, the invariant mass distributions of the $\JPsi\PagL$, $\JPsi\Pp$, and $\PagL \Pp$ two-body
combinations of the $\PBp \to \JPsi\PagL \Pp$ decay products are investigated.
To account for the event selection efficiency, each event is assigned a weight equal to $1/\epsilon(M(\Pp\PagL), \ctK)$, where $\epsilon(M(\PagL\Pp), \ctK)$ is obtained from simulation, as described in Section~\ref{EffSec}.
Background subtraction is performed using the \textit{sPlot} technique, with the $\JPsi\PagL\Pp$ invariant mass as the discriminating variable.
Figure~\ref{m12} shows the efficiency-corrected and background-subtracted invariant mass distributions of the $\JPsi\Pp$, $\JPsi\PagL$, and $\PagL \Pp$ systems.
These invariant mass distributions are compared with the pure phase space decay hypothesis (shown by the dashed lines), obtained from the generator-level simulation.
The data are poorly described by the pure phase space hypothesis in all three distributions. The degree of incompatibility between the data and this hypothesis is measured using the likelihood ratio method and is determined to be at least  6.1,  5.5, and 3.4 standard deviations for the $\JPsi\Pp$, $\JPsi\PagL$, and $\PagL \Pp$ invariant mass distributions, respectively. More details of the significance calculation are given in Section~\ref{significance}. We conclude that none of the three mass spectra can be adequately described by a pure three-body nonresonant phase space decay hypothesis, which is an indication of more complex dynamics in the $\PBp \to \JPsi\PagL \Pp$ decay.

There are at least three known $\Kstarp$ resonances, which we designate as $\Kstarp_{2,3,4}$, that can decay to $\PagL\Pp$, as listed in Table~\ref{Kstars}~\cite{PDG}. Even though the three $\Kstarp_{2,3,4}$ resonances listed in Table~\ref{Kstars} are beyond the kinematic region of the $\PBp \to \JPsi\PagL \Pp$ decay, these broad excited kaon states can contribute to the $\JPsi\Pp$ and $\JPsi\PagL$ invariant mass distributions, altering the pure phase space distributions.

To account for possible contributions from these resonances, we use a model-independent approach developed by BaBar in a search for the $\PZ(4430)$ resonance in the $\JPsi\Pgpp$ and $\psi(2S)\Pgpp$ channels~\cite{babarZ4430}. This approach was later used by LHCb in a similar search for the $\PZ(4430)$ particle in the $\psi(2S)\Pgpp$ invariant mass spectrum~\cite{LHCbZ4430} and to support the observation of possible pentaquark states $\PP_{\PQc}(4380)^{+}$ and $\PP_{\PQc}(4450)^{+}$ in the $\JPsi\Pp$ system~\cite{MIPc}. This method tests whether the contributions from a reflection of the resonant $\PagL\Pp$ angular amplitudes in the $\JPsi\PagL$ and $\JPsi\Pp$ spectra are sufficient to describe the data.

\begin{figure}[htb]
\centering
\includegraphics[width=0.49\textwidth]{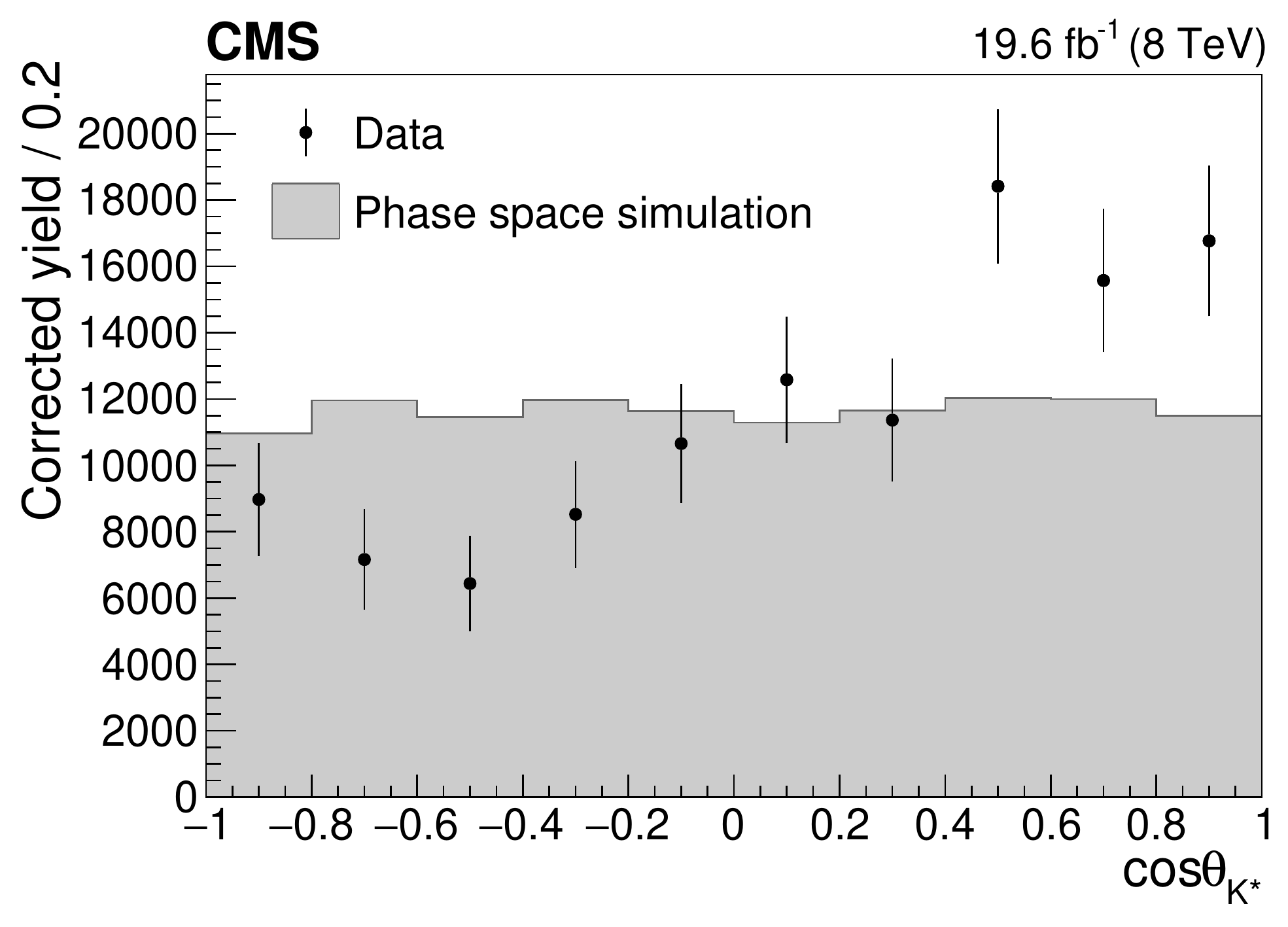}
\caption{
The background-subtracted and efficiency-corrected $\ctK$ distribution from the data (points with vertical bars) and the phase space simulation (shaded histogram). The vertical bars represent the statistical uncertainty.
}
\label{cosdata}
\end{figure}

The background-subtracted and efficiency-corrected $\ctK$ distribution in data is shown in Fig.~\ref{cosdata}. It is clear that, unlike the simulation of $\PBp \to \JPsi\PagL \Pp$ decay based on a pure phase space hypothesis, shown in the same figure, the distribution from data is not flat and has a structure that could affect the two-body invariant mass distributions under study.

In bins of $M(\PagL \Pp)$, the $\ctK$ distribution can be expressed as an expansion in terms of Legendre polynomials:
\begin{linenomath}
\begin{equation}\label{cosDistrPol}
\frac{\rd N}{ \rd\ctK} = \sum_{j=0}^{l_\text{max}}\langle P_{j}^{U}\rangle P_{j}(\ctK),
\end{equation}
\end{linenomath}
where $N$ is the efficiency-corrected and background-subtracted yield, $l_\text{max}$ depends on the maximum angular momentum used to describe the data, $P_{j}$ are the Legendre polynomials, and $\langle P_{j}^{U}\rangle$ are the unnormalized Legendre moments. The Legendre moments contain the full angular information of the $\PagL \Pp$ system and can be expressed using the following formula obtained by projecting the moments in Eq.~(\ref{cosDistrPol}) onto a Legendre polynomial basis:
\begin{linenomath}
\begin{equation}
\langle P_{j}^{U}\rangle = \sum_{i=1}^{N_\text{reco}}\frac{w^{i}_{\textit{sPlot}}}{\epsilon^{i}}P_{j}(\ctK),
\end{equation}
\end{linenomath}
where $N_\text{reco}$ is the number of selected events in each $M(\PagL\Pp)$ bin, $\epsilon^{i} = \epsilon^{i}(\cos\theta^{i}_{\kaon^{*}}, M(\PagL\Pp))$ is the efficiency correction factor, obtained as described in Section~\ref{EffSec}, and $w^{i}_\textit{sPlot}$ is the $\textit{sPlot}$ background subtraction weight.

\begin{figure}[htb!]
\centering
\includegraphics[width=0.4\textwidth]{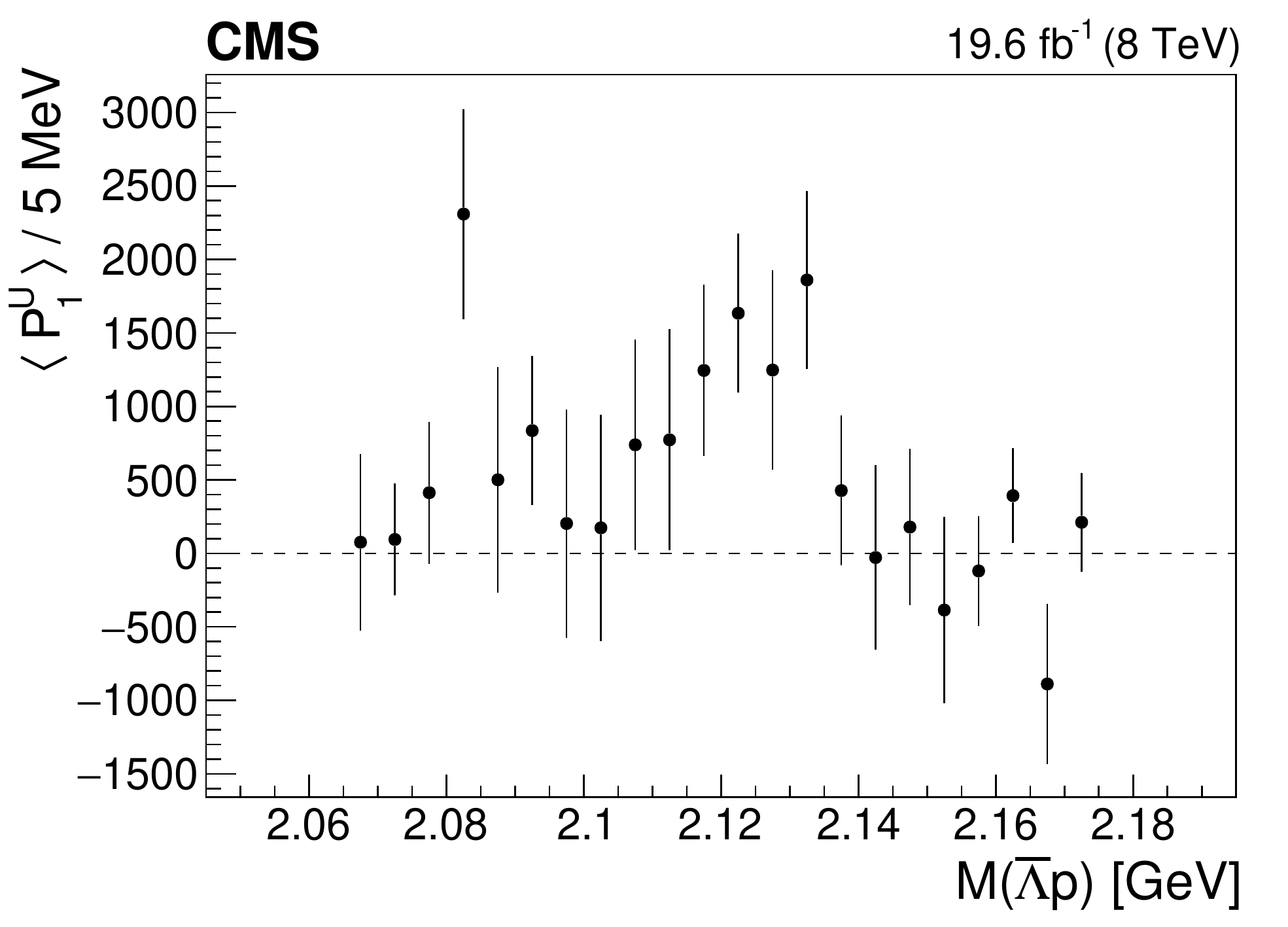}
\includegraphics[width=0.4\textwidth]{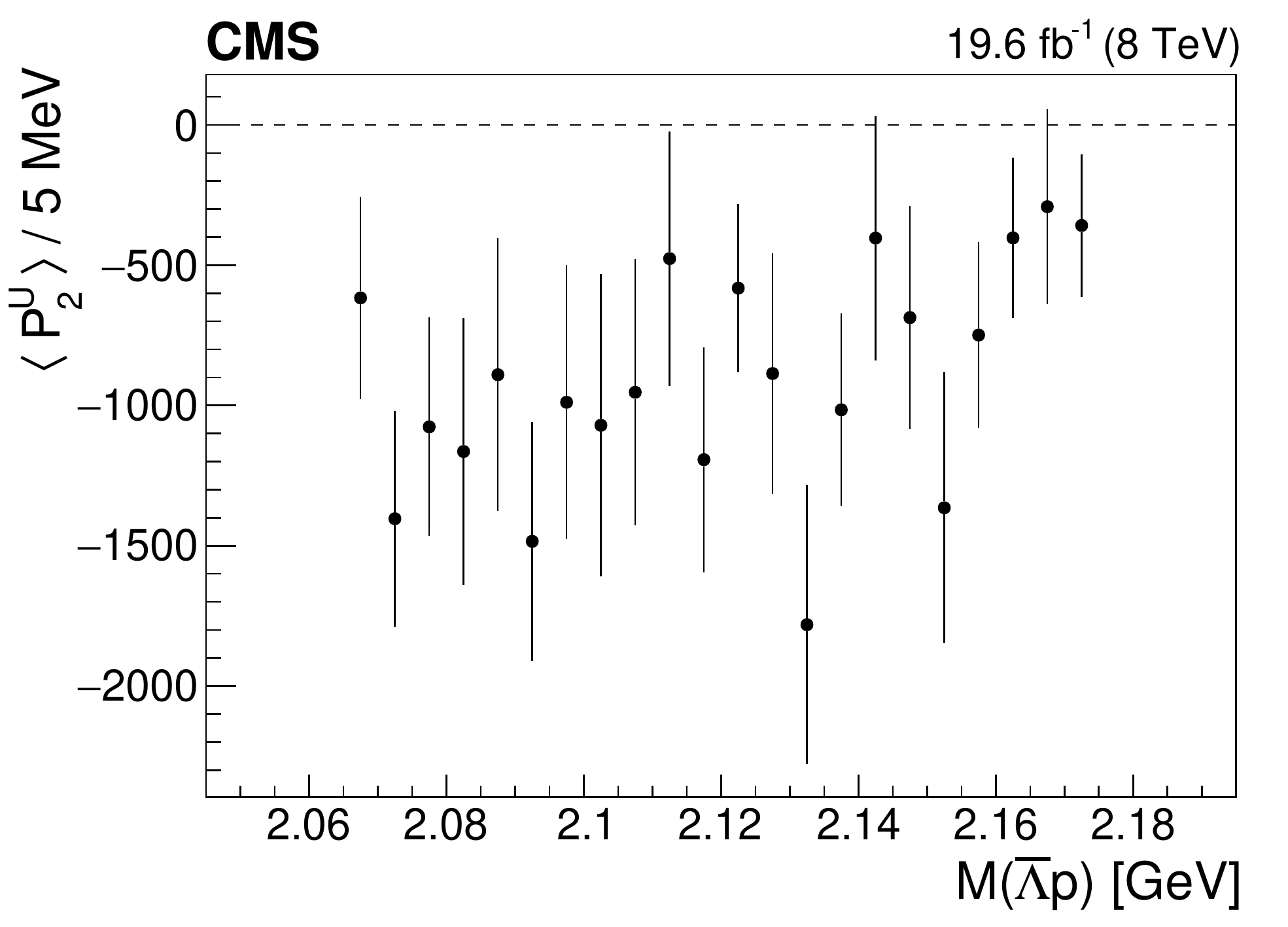}\\
\includegraphics[width=0.4\textwidth]{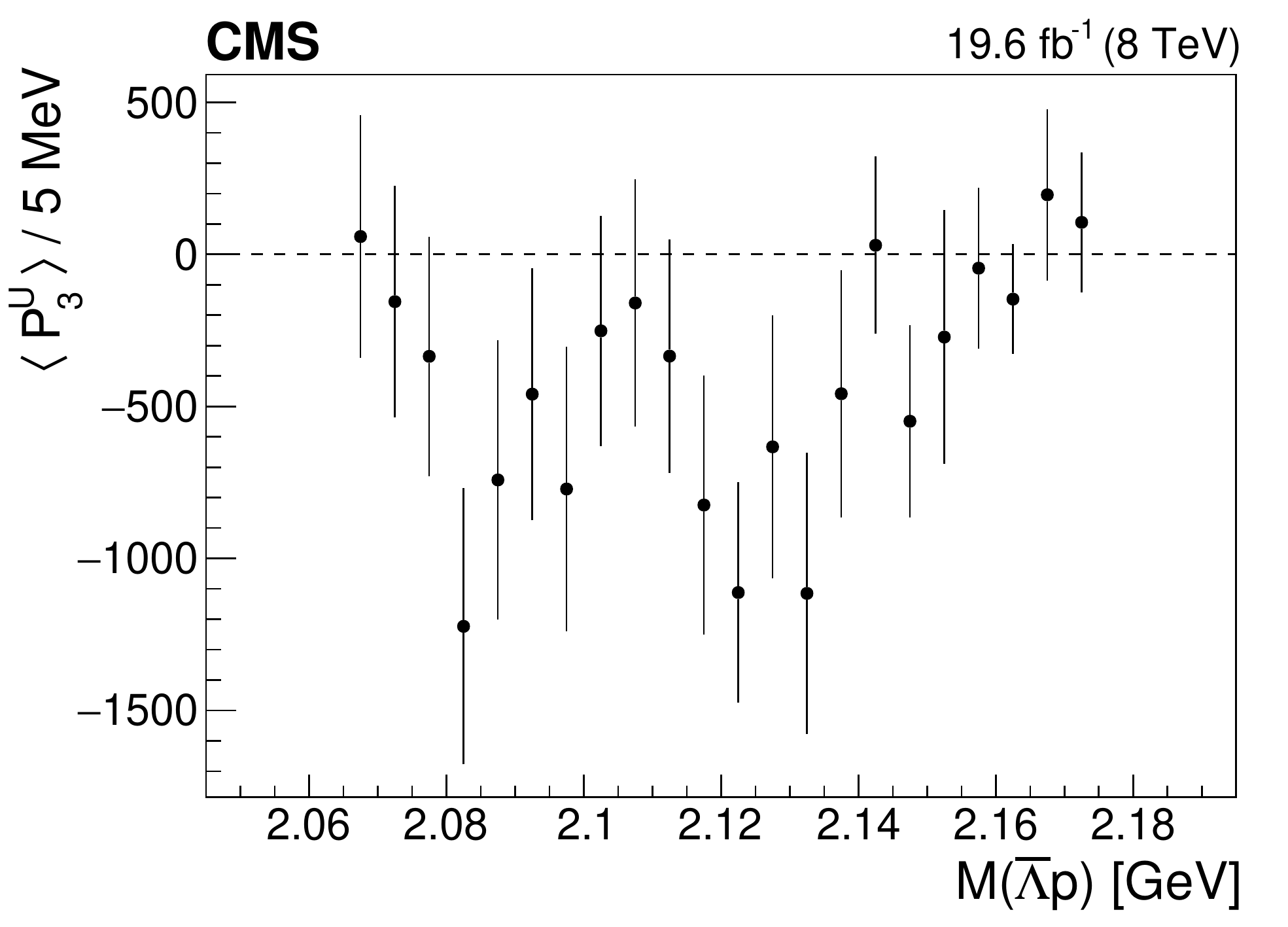}
\includegraphics[width=0.4\textwidth]{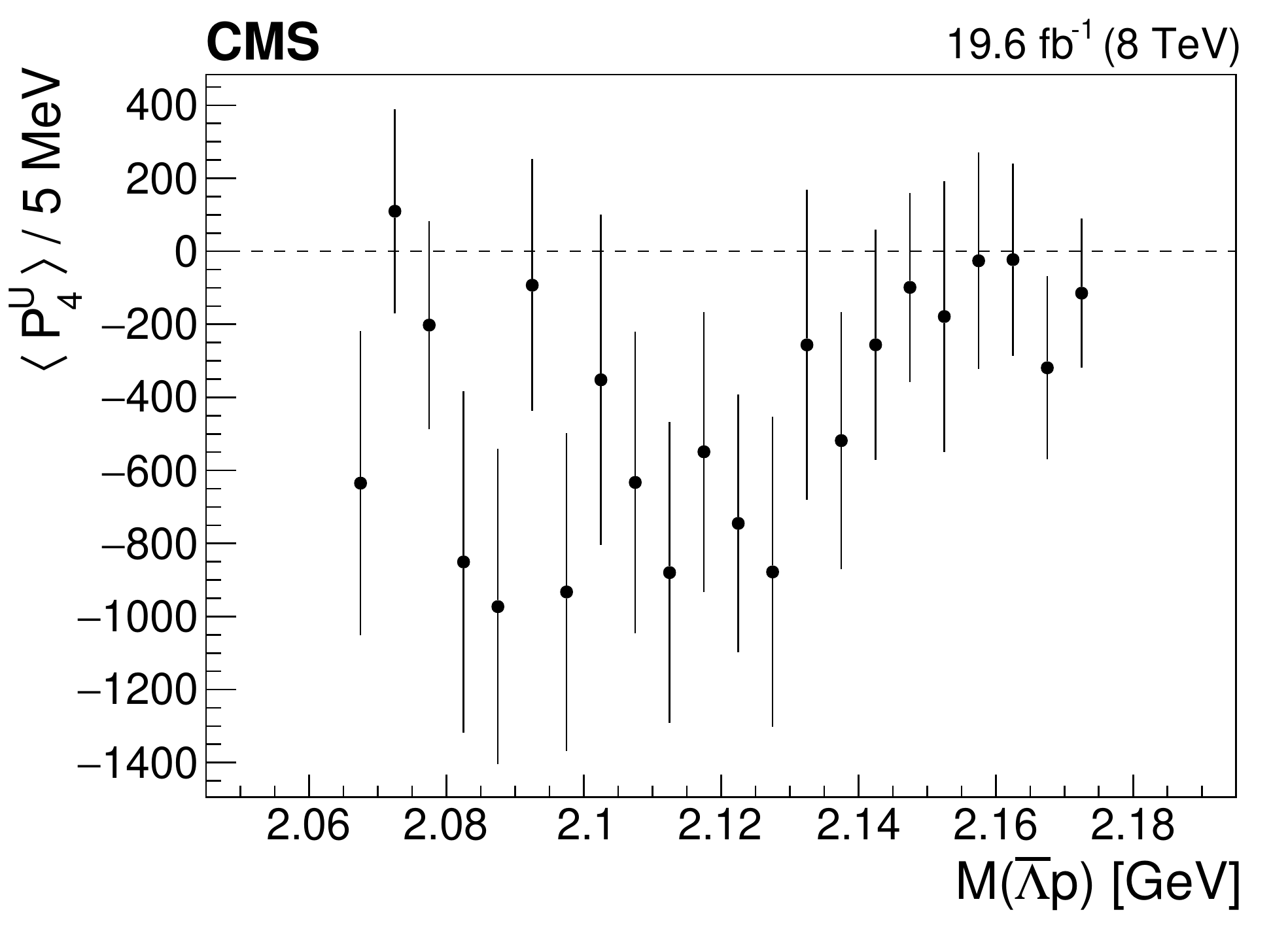}\\
\includegraphics[width=0.4\textwidth]{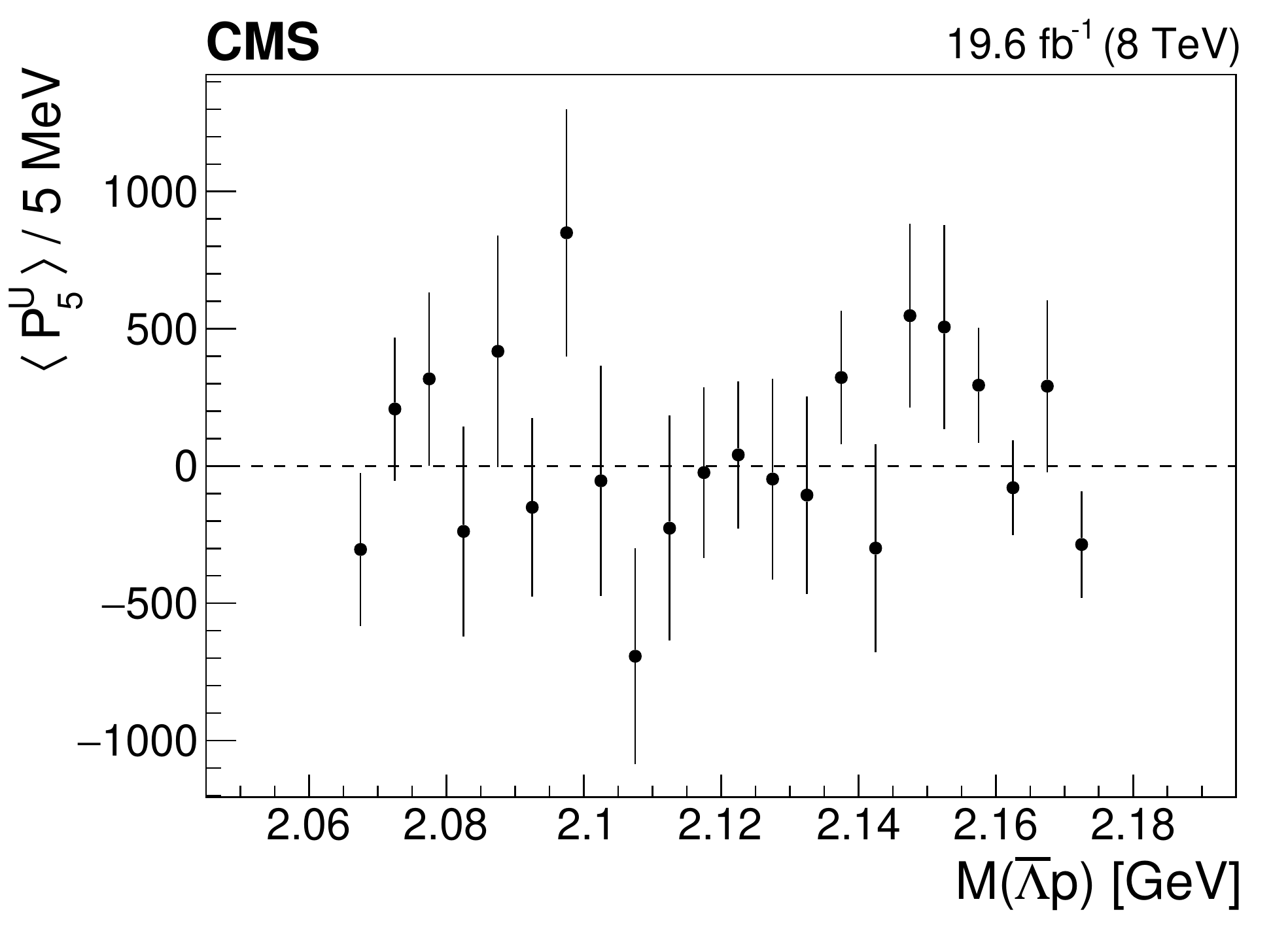}
\includegraphics[width=0.4\textwidth]{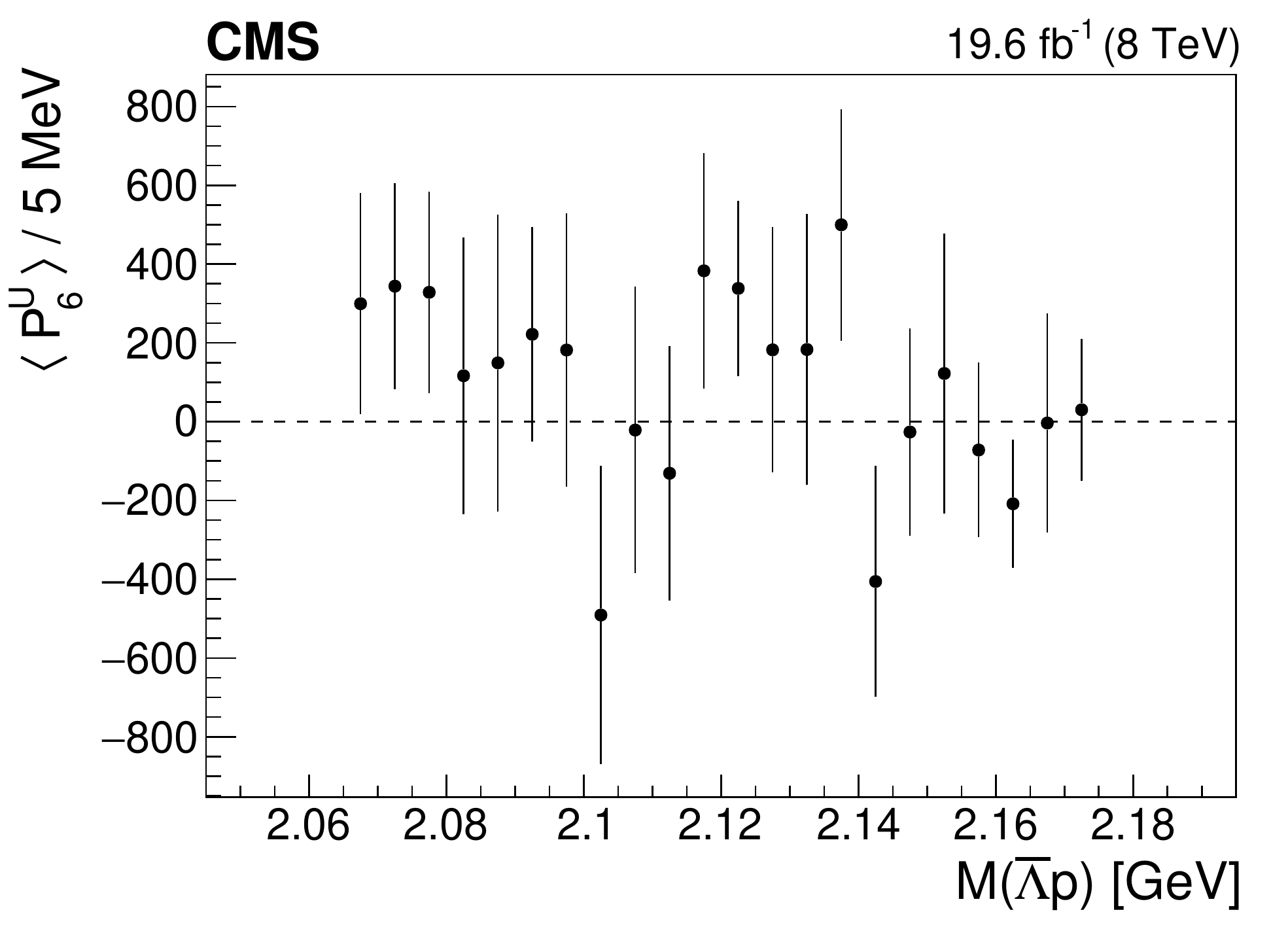}\\
\includegraphics[width=0.4\textwidth]{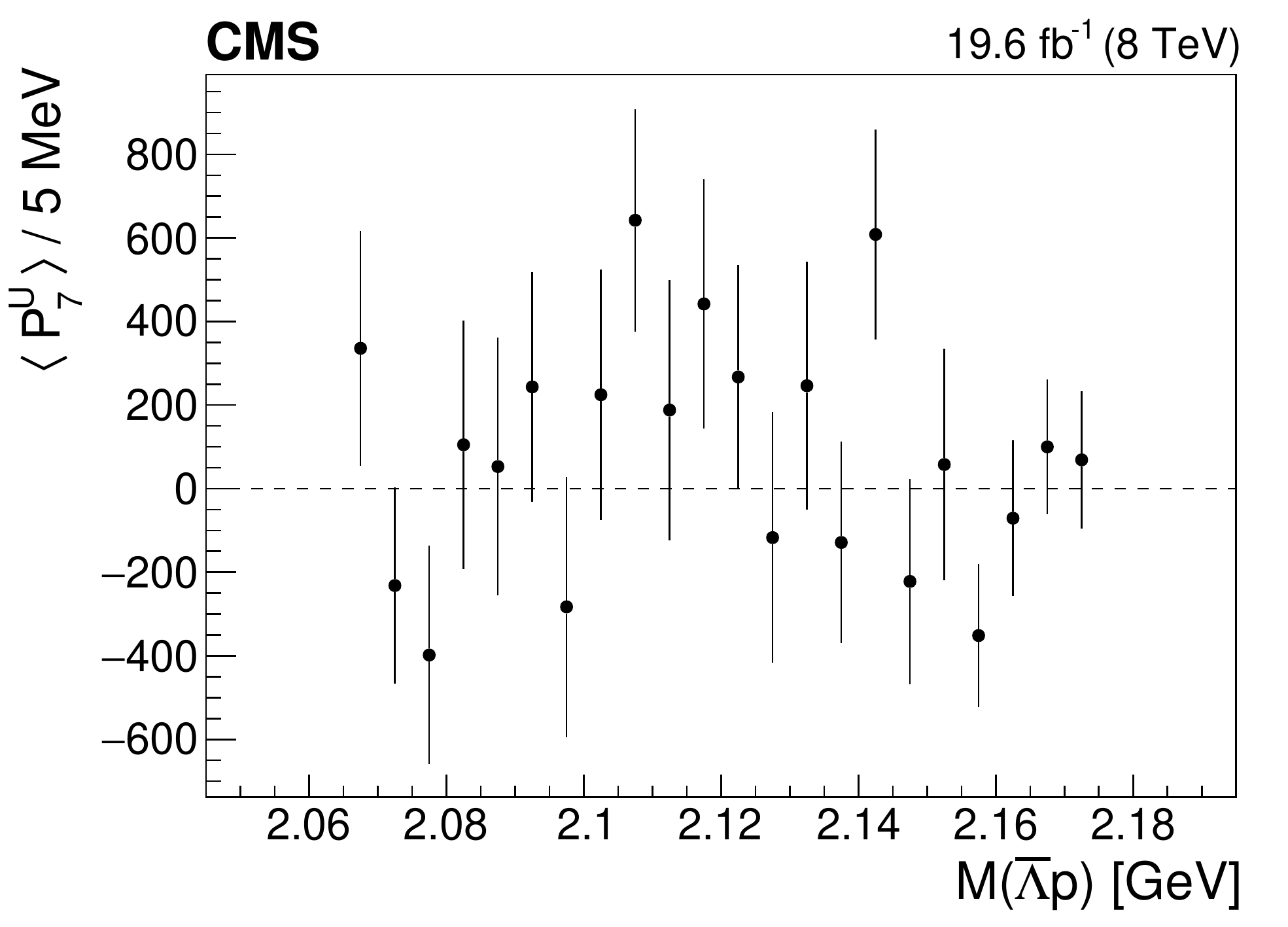}
\includegraphics[width=0.4\textwidth]{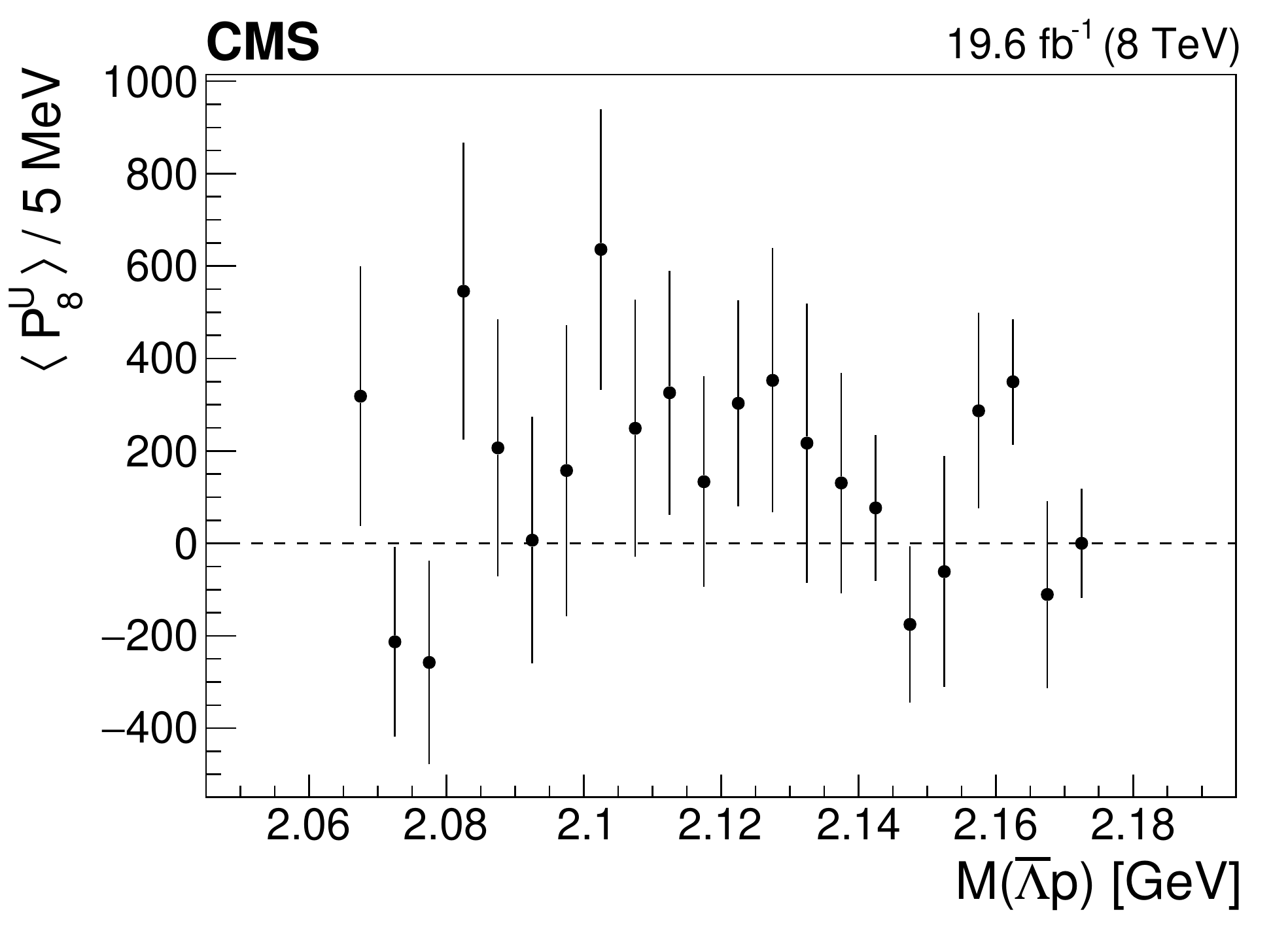}\\
\caption{
The dependence of the first eight Legendre moments on $M(\PagL\Pp)$. The vertical bars represent the statistical uncertainty.
}
\label{UnMoments}
\end{figure}

To determine the proper value for $l_\text{max}$, we note that the $\rd N/\rd\ctK$ distribution for a $\kaon^*$ resonance with spin $S$ is proportional to the square of the Wigner $d$ function $d^{S(\kaon^{*})}_{\lambda_{\kaon^{*}},(\lambda_{\Pp}-\lambda_{\PagL})}(\theta_{\kaon^{*}})$, given by the following equation:
\begin{linenomath}
\begin{equation}\label{D1}
d^{S}_{\lambda_{1}, \lambda_{2}}(\theta_{\kaon^{*}}) = \sum_{s=s_{\text{min}}}^{s_{\text{max}}} (-1)^s \frac{\sqrt{(S+\lambda_{2})!(S-\lambda_{2})!(S+\lambda_{1})!(S-\lambda_{1})!} }{(S+\lambda_{2}-s)!(S-\lambda_{1}-s)!(s-\lambda_{2}+\lambda_{1})!s! } \left(\cos\frac{\theta_{\kaon^{*}}}{2}\right)^{n_{1}}\left(\sin\frac{\theta_{\kaon^{*}}}{2}\right)^{n_{2}},
\end{equation}
\end{linenomath}
where $n_{1} =2S + \lambda_{2} - \lambda_{1} - 2s$, $n_{2} = \lambda_{1} - \lambda_{2} + 2s$, $\lambda_{1} = \lambda_{\kaon^{*}}$, and $\lambda_{2} = \lambda_{\Pp}-\lambda_{\PagL}$ are the corresponding spin projections, and $s$ is an integer such that:
\begin{linenomath}
\begin{equation}
s_{\text{min}} = \text{min}\{0,\lambda_{2} - \lambda_{1}\}\geq-2S,~s_{\text{max}} = \text{max}\{0,S+\lambda_{2},~S - \lambda_{1}\}\leq 2S.
\end{equation}
\end{linenomath}
By expanding Eq.~(\ref{D1}), one can see that the maximum power of $\ctK$ in the expression for $\rd N/\rd\ctK \sim  (d^{S}_{\lambda_{1}, \lambda_{2}}(\theta_{\kaon^{*}}))^{2}$ is given by $l_\text{max} = n_{1}+n_{2} = 2S$. Similarly, if one considers the interference between two resonances with spins $S_1$ and $S_2$, $l_\text{max} = S_1 + S_2$. Consequently, in order to fully describe the contributions from the resonances listed in Table~\ref{Kstars}, including their interference, it is sufficient to consider the Legendre moments up to twice the maximum spin of the considered resonances, i.e., $l_\text{max} = 8$.
The dependence of the first eight Legendre moments on $M(\PagL \Pp)$ from data is shown in Fig.~\ref{UnMoments}. If there were no resonant contributions to the intermediate two-body systems in the $\PBp \to \JPsi\PagL \Pp$ decay, the distributions shown in Fig.~\ref{UnMoments} would all be consistent with zero, which is not the case.

To investigate whether the $\PagL \Pp$ angular structure caused by the $\Kstarp_{2,3,4}$ resonances with spins up to four is sufficient to describe the data, a reweighting of the simulated signal sample is performed and the result is compared with the background-subtracted and efficiency-corrected data. A significant disagreement between the data and the reweighted simulation that accounts for the invariant mass of the $\PagL\Pp$ system and the angular structure corresponding to the resonances with the spin up to four in the $\PagL\Pp$ system may indicate the presence of an exotic state in the $\JPsi\PagL$ or $\JPsi\Pp$ systems. Since the data are corrected for the reconstruction efficiency, we use generator-level simulated samples without detector simulation for the reweighting. The simulation is forced to reproduce the $\PagL\Pp$ invariant mass spectrum observed in data by applying a weight to each simulated event obtained using a linear interpolation of the data-to-simulation ratio histogram. The angular structure is introduced into the simulation by using appropriate weights, described below.
To obtain the weights corresponding to the angular structure, Eq.~(\ref{cosDistrPol}) is expanded as follows:
\begin{linenomath}
\begin{equation}\label{weights0}
\frac{ \rd N}{ \rd\ctK} = \frac{N}{2} + \sum_{j=1}^{l_\text{max}}\langle P_{j}^{U}\rangle P_{j}(\ctK) = \frac{N}{2}(1 + \sum_{j=1}^{l_\text{max}}\langle \frac{ 2\langle P_{j}^{U}\rangle}{N}\rangle P_{j}(\ctK)).\\
\end{equation}
\end{linenomath}
The factor of $N/2$ in Eq.~(\ref{weights0}) appears because of the Legendre polynomial normalization convention, i.e.,
 $\langle P_{0}^{N}\rangle P_{0}(\ctK) =1/2$. By factoring out the $N/2$ term from the right-hand side of Eq.~(\ref{weights0}), one can obtain the weights given by the following equation:
\begin{linenomath}
\begin{equation}\label{weights}
w^{i} =1 + \sum_{j=1}^{l_\text{max}}\langle P_{j}^{N}\rangle P_{j}(\ctK),
\end{equation}
\end{linenomath}
where $\langle P_{j}^{N}\rangle =2\langle P_{j}^{U}\rangle/N_\text{reco}^\text{corr}$ are the normalized Legendre moments and $N_\text{reco}^\text{corr}$ is the corrected number of reconstructed events in each $M(\PagL\Pp)$ bin.

The templates obtained after applying the weights corresponding to the observed $\PagL \Pp$ invariant mass structure, as well as the weights given by Eq.~(\ref{weights}) applied to the simulation, are then compared with the efficiency-corrected and background-subtracted data. Results are shown in Fig.~\ref{m12} (solid line). As seen from the lower panel in the figure, the $M(\PagL \Pp)$ distribution in data is well described by the reweighted simulation, which is expected by the construction of the weights. The $M(\PagL \Pp)$ distribution is not affected by the $w^{i}$ weights since the integrals of the individual $P_{j}(x)$ functions over the full range in $\ctK$ are equal to zero. It is also evident from the two upper panels in Fig.~\ref{m12} that the description of the $M(\JPsi\Pp)$ and $M(\JPsi \PagL)$ data distributions is improved after accounting for both the observed angular and invariant mass structures in the $\PagL \Pp$ system.

\subsection{Significance calculation}\label{significance}

In this section, the compatibility of the data with both hypotheses of pure phase space ($H_{\text{PS}}$) and phase space augmented with the eight Legendre moments and the reweighting of the $M(\PagL \Pp)$ distribution to describe the structure observed in data ($H_{\text{L8}}$) is quantified using the likelihood ratio technique. To test the compatibility of data with the $H_{\text{L8}}$ hypothesis, 2000 pseudo-experiments were generated, each with the number of signal events, $N$, equal to the one observed in data, according to the probability density function corresponding to this hypothesis $F^{X}(H_{\text{L8}})$, where $X$ stands for the projection on the invariant mass of the corresponding system.
An additional hypothesis $H_{\cos\theta}$ has a probability density function $F^{X}(H_{\cos\theta})$ that accounts for all the features in the $\ctK$ and $M(\PagL\Pp)$ distributions observed in data. It is obtained by reweighting the pure phase space simulation to reproduce  the $\ctK$ distribution in data (shown in Fig.~\ref{cosdata}) in each $M(\PagL\Pp)$ bin, as well as by reweighting the $M(\PagL\Pp)$ spectrum in simulation to match the one observed in data. Therefore, the $F^{X}(H_{\cos\theta})$ function reflects the total angular structure of the $\PagL\Pp$ system and provides the best description of the $M(\JPsi \PagL)$ and $M(\JPsi\Pp)$ invariant mass spectra. The logarithm of the likelihood ratio is used to define the test statistic:
\begin{linenomath}
\begin{equation}
2\Delta \text{NLL} = -2\sum_{i=1}^{N}\ln\frac{F^{X}(H_{\text{L8}})}{F^{X}(H_{\cos\theta})},
\end{equation}
\end{linenomath}
where the sum runs over the events in each pseudo-experiment or in data.

The $2\Delta \text{NLL}$ distribution from the pseudo-experiments is well described by a Gaussian function; the $2\Delta \text{NLL}_\text{data}$ value is calculated using collision data. The significance of the $H_{\text{L8}}$ hypothesis incompatibility with data is calculated as the number of standard deviations between the observed $2\Delta \text{NLL}_\text{data}$ value and the mean value of the $2\Delta \text{NLL}$ distribution from the pseudo-experiments.

The same test is performed to quantify the incompatibility of the data with the pure phase space hypothesis $H_{\text{PS}}$, following the procedure described for the test of $H_{\text{L8}}$.

There are several sources of systematic uncertainty  that could affect the significance calculation. The first source is the function used to describe the background component in the  $M(\JPsi \PagL \Pp)$ distribution, which enters through the \textit{sPlot} background-subtraction procedure. This uncertainty is estimated by using two alternative models for the background component in the $M(\JPsi \PagL \Pp)$ spectrum: the baseline model multiplied by a polynomial of either first or second order. Another source of systematic uncertainty is due to statistical fluctuations in the 2D efficiency calculation, discussed in Section~\ref{EffSec}. To test how the significance is affected by these fluctuations, additional parameterizations of the 2D efficiency are considered: a histogram with wider bins and a fit to the efficiency distribution with 2D polynomials, with the uncertainties obtained from the fit taken into account.

The contribution of the systematic uncertainty to the significance calculation from the kinematic requirements used in the selection of the $\PBp \to \JPsi\PagL \Pp$ candidates is also estimated.
The effect of the requirement applied to the reflected invariant mass of the $\PagL$ candidate daughters $\abs{M(\Pgpp\Pgpm)-M_{\PKzS}^\text{PDG}}>2\sigma^{\PKzS}_\text{eff}$ is evaluated by repeating the significance calculation without this requirement. The effect of the $\pt$ selection criteria applied to the $\PBp$ meson and the $\JPsi$, $\PagL$, and $\Pp$ candidates in the final state of the $\PBp \to \JPsi\PagL \Pp$ decay, was also tested by tightening the baseline requirements by 50$\%$ and recalculating the significance values.

The effect of the correlation between the $M(\PagL\Pp)$, $\ctK$, $\JPsi\Pp$, and $\JPsi \PagL$ variables is tested by generating pseudo-experiments according to the two-dimensional probability density function $F^{Y}(H_{\text{L8}})$, where $\text{Y} = (M(\PagL\Pp),~\ctK)$, and then projecting them to the $\JPsi\Pp$ system invariant mass. The significance values decrease on average by 30\%, which does not exceed the significance range introduced by the effects of the other systematics sources discussed above.

Under the variations discussed above, the significance of the incompatibility of data with the $H_{\text{PS}}$ is found to vary from 6.1 to 8.1, 5.5 to 7.4, and 3.4 to 4.8 standard deviations for the $\JPsi\Pp$, $\JPsi \PagL$, and $\PagL\Pp$ invariant mass distributions, respectively. The incompatibility of data with the phase space augmented with the eight Legendre moments and the reweighting of the $M(\PagL \Pp)$ distribution to describe the structure observed in data $H_{\text{L8}}$ varies from 1.3 to 2.8 (2.7) standard deviations for the $\JPsi\Pp$ ($\JPsi \PagL$) invariant mass spectrum, which allows us to conclude that the data are consistent with the $H_{\text{L8}}$ hypothesis. We note that since the quoted statistical significances for the $\JPsi\Pp$ and $\JPsi \PagL$ systems are not independent, they cannot be combined without properly taking into account the correlation. While the presence of new resonances in an intermediate two-body system produced in this decay cannot be ruled out, the deviation of the data from a pure phase space model can be adequately described by including only the known $\Kstarp_{2,3,4}$ resonances in the $\PagL\Pp$ system.

\section{Summary}
{\tolerance=800
Using a data set of proton-proton collisions collected by the CMS experiment at $\sqrt{s}=8\TeV$ and corresponding to an integrated luminosity of 19.6\fbinv, the ratio of branching fractions has been measured to be ${\cal B}(\PBp \to \JPsi\PagL \Pp)/{\cal B}(\PBp \to \JPsi \Kstarpz) = (1.054\pm0.057\stat \pm0.035\syst\pm0.011({\cal B}))\%$. Using the world-average branching fraction of  the $\PBp \to \JPsi \Kstarpz$ decay, the branching fraction of the $\PBp \to \JPsi\PagL \Pp$ decay is determined to be $(15.1\pm0.8\stat \pm0.5\syst\pm0.9({\cal B}))\times 10^{-6}$, the most precise measurement to date.
A study of the two-body invariant mass distributions of the $\PBp \to \JPsi\PagL \Pp$ decay products demonstrates that these spectra cannot be adequately modeled with a pure phase space decay hypothesis. The incompatibility of the data with this hypothesis is more than  6.1, 5.5, and 3.4 standard deviations for the $\JPsi\Pp$, $\JPsi \PagL$, and $\PagL \Pp$ invariant mass spectra, respectively. A model-independent approach that accounts for the contribution from known $\Kstarp_{2,3,4}$ resonances with spins up to 4 decaying to the $\PagL \Pp$ system improves the agreement significantly, decreasing the incompatibility with data to less than three standard deviations in both the $\JPsi\Pp$ and $\JPsi \PagL$ invariant mass spectra.
\par}
\begin{acknowledgments}
We congratulate our colleagues in the CERN accelerator departments for the excellent performance of the LHC and thank the technical and administrative staffs at CERN and at other CMS institutes for their contributions to the success of the CMS effort. In addition, we gratefully acknowledge the computing centers and personnel of the Worldwide LHC Computing Grid for delivering so effectively the computing infrastructure essential to our analyses. Finally, we acknowledge the enduring support for the construction and operation of the LHC and the CMS detector provided by the following funding agencies: BMBWF and FWF (Austria); FNRS and FWO (Belgium); CNPq, CAPES, FAPERJ, FAPERGS, and FAPESP (Brazil); MES (Bulgaria); CERN; CAS, MoST, and NSFC (China); COLCIENCIAS (Colombia); MSES and CSF (Croatia); RPF (Cyprus); SENESCYT (Ecuador); MoER, ERC IUT, and ERDF (Estonia); Academy of Finland, MEC, and HIP (Finland); CEA and CNRS/IN2P3 (France); BMBF, DFG, and HGF (Germany); GSRT (Greece); NKFIA (Hungary); DAE and DST (India); IPM (Iran); SFI (Ireland); INFN (Italy); MSIP and NRF (Republic of Korea); MES (Latvia); LAS (Lithuania); MOE and UM (Malaysia); BUAP, CINVESTAV, CONACYT, LNS, SEP, and UASLP-FAI (Mexico); MOS (Montenegro); MBIE (New Zealand); PAEC (Pakistan); MSHE and NSC (Poland); FCT (Portugal); JINR (Dubna); MON, RosAtom, RAS, RFBR, and NRC KI (Russia); MESTD (Serbia); SEIDI, CPAN, PCTI, and FEDER (Spain); MOSTR (Sri Lanka); Swiss Funding Agencies (Switzerland); MST (Taipei); ThEPCenter, IPST, STAR, and NSTDA (Thailand); TUBITAK and TAEK (Turkey); NASU and SFFR (Ukraine); STFC (United Kingdom); DOE and NSF (USA).

\hyphenation{Rachada-pisek} Individuals have received support from the Marie-Curie program and the European Research Council and Horizon 2020 Grant, contract Nos.\ 675440 and 765710 (European Union); the Leventis Foundation; the A.P.\ Sloan Foundation; the Alexander von Humboldt Foundation; the Belgian Federal Science Policy Office; the Fonds pour la Formation \`a la Recherche dans l'Industrie et dans l'Agriculture (FRIA-Belgium); the Agentschap voor Innovatie door Wetenschap en Technologie (IWT-Belgium); the F.R.S.-FNRS and FWO (Belgium) under the ``Excellence of Science -- EOS" -- be.h project n.\ 30820817; the Beijing Municipal Science \& Technology Commission, No. Z181100004218003; the Ministry of Education, Youth and Sports (MEYS) of the Czech Republic; the Lend\"ulet (``Momentum") Programme and the J\'anos Bolyai Research Scholarship of the Hungarian Academy of Sciences, the New National Excellence Program \'UNKP, the NKFIA research grants 123842, 123959, 124845, 124850, 125105, 128713, 128786, and 129058 (Hungary); the Council of Science and Industrial Research, India; the HOMING PLUS program of the Foundation for Polish Science, cofinanced from European Union, Regional Development Fund, the Mobility Plus program of the Ministry of Science and Higher Education, the National Science Center (Poland), contracts Harmonia 2014/14/M/ST2/00428, Opus 2014/13/B/ST2/02543, 2014/15/B/ST2/03998, and 2015/19/B/ST2/02861, Sonata-bis 2012/07/E/ST2/01406; the Ministry of Education and Science of the Russian Federation contract No. 14.W03.31.0026; the National Priorities Research Program by Qatar National Research Fund; the Programa Estatal de Fomento de la Investigaci{\'o}n Cient{\'i}fica y T{\'e}cnica de Excelencia Mar\'{\i}a de Maeztu, grant MDM-2015-0509 and the Programa Severo Ochoa del Principado de Asturias; the Thalis and Aristeia programs cofinanced by EU-ESF and the Greek NSRF; the Rachadapisek Sompot Fund for Postdoctoral Fellowship, Chulalongkorn University and the Chulalongkorn Academic into Its 2nd Century Project Advancement Project (Thailand); the Welch Foundation, contract C-1845; and the Weston Havens Foundation (USA).
\end{acknowledgments}

\bibliography{auto_generated}

\cleardoublepage \appendix\section{The CMS Collaboration \label{app:collab}}\begin{sloppypar}\hyphenpenalty=5000\widowpenalty=500\clubpenalty=5000\input{BPH-18-005-authorlist.tex}\end{sloppypar}
\end{document}

%% file: BPH-18-005-authorlist.tex
\vskip\cmsinstskip
\textbf{Yerevan Physics Institute, Yerevan, Armenia}\\*[0pt]
A.M.~Sirunyan$^{\textrm{\dag}}$, A.~Tumasyan
\vskip\cmsinstskip
\textbf{Institut f\"{u}r Hochenergiephysik, Wien, Austria}\\*[0pt]
W.~Adam, F.~Ambrogi, T.~Bergauer, J.~Brandstetter, M.~Dragicevic, J.~Er\"{o}, A.~Escalante~Del~Valle, M.~Flechl, R.~Fr\"{u}hwirth\cmsAuthorMark{1}, M.~Jeitler\cmsAuthorMark{1}, N.~Krammer, I.~Kr\"{a}tschmer, D.~Liko, T.~Madlener, I.~Mikulec, N.~Rad, J.~Schieck\cmsAuthorMark{1}, R.~Sch\"{o}fbeck, M.~Spanring, D.~Spitzbart, W.~Waltenberger, C.-E.~Wulz\cmsAuthorMark{1}, M.~Zarucki
% ...
\vskip\cmsinstskip
\dag: Deceased\\
1:  Also at Vienna University of Technology, Vienna, Austria\\
% ...
